\documentclass[reqno,12pt]{amsart}

\usepackage{enumerate}
\usepackage[margin=1in]{geometry}
\usepackage{enumitem}
\usepackage{ifpdf}
\usepackage{amsmath}
\usepackage{amsfonts}
\usepackage{amssymb}
\usepackage{amsaddr}
\usepackage{amsthm}
\usepackage{mathrsfs}
\usepackage{amsrefs}
\usepackage{amsfonts}
\usepackage{amssymb}
\usepackage{cite}
\usepackage[hyperfootnotes=false]{hyperref}
\usepackage[titletoc,title]{appendix}
\usepackage[dotinlabels]{titletoc}
\usepackage[all,cmtip]{xy}
\usepackage{nicefrac}
\usepackage{float}
\usepackage{verbatim}
\usepackage[multiple]{footmisc}
\usepackage{mathdots}
\usepackage{dcolumn}
\usepackage{tabu}
\usepackage[hang,small,bf]{caption}    

\usepackage{tikz}	
\usetikzlibrary{backgrounds,fit,decorations.pathreplacing}  

\newlength\Li \newlength\Lii 
\newcommand{\ignore}[1]{}

\newcommand{\abs}[1]{\left\lvert {#1} \right\rvert}
\newcommand{\norm}[1]{\left\lVert {#1} \right\rVert}

\theoremstyle{definition}

\theoremstyle{remark}

\theoremstyle{note}

\usepackage{tikz}
\usetikzlibrary{backgrounds,fit,calc,trees,positioning,arrows,chains,shapes.geometric,%
    decorations.pathreplacing,decorations.pathmorphing,shapes,%
    matrix,shapes.symbols,snakes}

\newcommand{\ket}[1]{\ensuremath{\left|#1\right\rangle}} 

\tikzstyle{sum} = [draw,circle,inner sep=0mm,minimum size=2mm]
\tikzstyle{connector} = [->,thick]
\tikzstyle{line} = [thick]
\tikzstyle{branch} = [circle,inner sep=0pt,minimum size=1mm,fill=black,draw=black]
\tikzstyle{guide} = []
\tikzstyle{snakeline} = [connector, decorate, decoration={pre length=0.2cm,
                         post length=0.2cm, snake, amplitude=.4mm,
                         segment length=2mm},thick, magenta, ->]

\makeatletter
\newcommand*{\rom}[1]{\expandafter\@slowromancap\romannumeral #1@}
\makeatother

\newtheoremstyle{defnopunct}
{3pt}
{3pt}
{}
{}
{\bfseries}
{ }
{.5em}
{}

\usepackage{etoolbox}
\let\bbordermatrix\bordermatrix
\patchcmd{\bbordermatrix}{8.75}{4.75}{}{}
\patchcmd{\bbordermatrix}{\left(}{\left[}{}{}
\patchcmd{\bbordermatrix}{\right)}{\right]}{}{}

\author{Asif Shakeel}
\email{asif.shakeel@gmail.com}
\ifpdf
\fi

\title[Efficient and scalable quantum walk algorithms] {Efficient and scalable quantum walk algorithms   \\   via the quantum Fourier transform}

\begin{document}

\begin{abstract}
Quantum walks (QWs) are of interest as  examples of uniquely quantum behavior and are applicable in a variety of quantum search and simulation models. Implementing  QWs on quantum devices is useful from both points of view. We describe a prototype   one-dimensional  discrete time QW algorithm  that economizes resources required in its implementation. Our algorithm  needs only a single   shift (increment) operation.  It also allows complete flexibility in choosing the shift circuit, a  resource intensive part of QW implementations.  We implement the shift using the quantum Fourier transform (QFT), yielding, to date,  the most efficient and scalable, quadratic size, linear depth circuit for the basic QW. This is desirable for   Noisy Intermediate-Scale Quantum   (NISQ) devices, in which  fewer  computations implies faster execution and reduced effects of noise and  decoherence.  As the QFT diagonalizes unitary circulant matrices, we generalize the shift in the basic QW to introduce   spatial  convolutions in the QW. We demonstrate  our basic QW algorithm using the QFT based shift by running it  on publicly accessible IBM  quantum computers. 

\end{abstract}

\maketitle

 \section{Introduction} \label{sec:intro}
Quantum walks (QWs) exhibit  properties  that are fundamentally quantum mechanical, serving to demonstrate in their most basic form, walk patterns that are unachievable by classical random walks. In their  sophisticated configurations, they appear in  applications in areas as widely apart as quantum simulation models~\cite{adma:deqwhctl, BerryQIC12} and  quantum search algorithms~\cite{mnrs:svqw}.    QWs have been experimentally realized on different physical systems~\cite{scpg:pwl,bflkaw:dspqwtd}.  The reader is referred to~\cite{sva:qwco} for a comprehensive review of QWs.  With the public availability of  quantum computers, quantum algorithms can be tested and their resource  requirements and performance examined. In the current era of Noisy Intermediate-Scale Quantum   (NISQ)~\cite{p:qcnisq} devices,  decoherence and noise may severely degrade  the  performance of a quantum algorithm implemented on a device, negating any  gains to be had from  a using a quantum algorithm over a classical one. As a consequence, the issue of  resource use in  implementation deserves  consideration.  In this paper we develop  a  computational description  of a one-dimensional discrete time QW that turns into a flexible, scalable and highly efficient implementation in terms of circuit size and depth.~\footnote{The circuit size is  the number of at most two-qubit gates used in the circuit. The circuit depth is the longest path length in gates from an input to an output.} It also illustrates that  partial optimizations of an algorithm can be done  mathematically  prior to circuit implementation on a quantum device

A basic discrete time QW  is specified by a particle walking on a lattice, at each time step moving from its current {\it site} position to another in its vicinity, and scattering  through {\it self-interaction}. We will, for this paper, consider  a one dimensional walk on a finite lattice.   The state of the QW is  specified by its  position and velocity. Taking the lattice to be of size $N$, the particle's position  $x \in \{0,\ldots,N-1\}$ then labels a basis element $\ket{x}$ of an $N$-dimensional {\it position}  Hilbert space.  At each time step, the particle can move one lattice position  to the ``left" $x \mapsto x-1$, or  to the ``right", $x \mapsto x+1$, where the addition/subtraction is modulo $N$.  This directional motion is encoded in a  two dimensional {\it velocity} Hilbert space with basis elements labeled  $ \ket{v}$,  $v =  -1,+1$, for left and right moving particle respectively. The QW Hilbert space $\mathcal{H}$ is given as
\begin{equation*}
\mathcal{H} = \mathbb{C}^2  \otimes \mathbb{C}^N
\end{equation*}
with  basis elements labeled $\ket{v,x}:=\ket{v}\otimes \ket{x}$.
 The state of a QW is an element $\psi \in \mathcal{H}$ of unit norm $\norm{\psi} = 1$.
A QW  evolves through two consecutive unitary actions on its state.
 \begin{enumerate}[label=(\roman{*})] 
\item \label{scat}  Scattering operation that acts on the velocity space, i.e., is of the form $S \otimes \mathbb{I}_x$,  
where $S$ is a unitary matrix on $\mathbb{C}^2$, the velocity space, typically~\cite{bib:meyer2},
\begin{equation*}
S=\begin{bmatrix}
ie^{i\alpha}\sin\theta&e^{i\alpha}\cos\theta\\
e^{i\alpha}\cos\theta&ie^{i\alpha}\sin\theta
\end{bmatrix}.
\end{equation*}  
and $\mathbb{I}_x$ is the   identity operator on the position space.
\item \label{prop} Propagation  operation that moves the particle in the direction of the velocity,
\begin{equation*}
\sigma : \ket{v,x}    \mapsto \ket{v,x+v},
\end{equation*}
where the addition is modulo $N$.
\end{enumerate}
Denoting 
\begin{equation} \label{hats}
\hat S = S \otimes  \mathbb{I}_x,
\end{equation}
a  step $T$ of QW evolution  is thus: 
\begin{equation*}
T = \sigma  \hat S.
\end{equation*}
This is the model of QW that we are going to be concerned with. 

The rest of this  paper is organized as follows.  In Section~\ref{sec:algQW} we describe our QW algorithm   that, for $n$-qubit position space, implements the right/left shifts (increment/decrement) in propagation $\sigma$ by  single right shift (increment) and $2n$ CNOT gates that control the direction.   We recall the  quantum Fourier transform (QFT) and compare  the shift implementation using the QFT  with the one based on generalized CNOT gates,   common in applications. In Section~\ref{sec:qftconv} we   introduce the QW whose  state, during propagation,   evolves through  spatial convolution, generalizing the shift. We also describe  a  strategy for efficient implementation of this family of QWs. In Section~\ref{sec:ibmsim} we run our basic QW algorithm with QFT based shift on some of the IBM quantum computers for a few lattice sizes and number of steps of evolution, and examine their performances. Section~\ref{sec:conc} is the conclusion.



\section{An algorithm for QW} \label{sec:algQW}

Let us first express the vectors and operators in the natural bases for the position and velocity  spaces. Considering the position space by itself, we can use   the natural ordering induced on the basis by values of $x$  and write $\ket{x}$ as a coordinate vector 
\begin{equation} \label{poseq}
\left| x \right \rangle= 
  \begin{bmatrix}
	0 \\
 	0\\
    \vdots\\
   	\vdots\\
   	0\\
   	1\\
	0\\
    \vdots\\
    0   
  \end{bmatrix}  
  \begin{array}{l}
	(0) \\
 	(1) \\
    \vdots \\
    \vdots \\
   	\phantom{0} \\
   	(x)\\
	\phantom{0} \\
    \vdots \\
    \phantom{0} 
    \end{array}
\end{equation}  
where the coordinate   indices are displayed in parentheses and all the component values are $0$, except at the index $x$,   whose component  value is $1$. Similarly, for the joint velocity-position space, we write $\ket{v,x}$ as a  coordinate vector of length $2N$, with the first $N$ elements representing  positions corresponding to $v=0$ and the last $N$, those corresponding  to  $v=1$, 
\begin{align*}
\ket{v,x}= 
  \begin{bmatrix}
	0 \\
 	0\\
    \vdots\\
    \vdots\\
   	0\\
   	1\\
	0\\
    \vdots\\
    0   
  \end{bmatrix}  
  \begin{array}{l}
	(0) \\
 	(1) \\
    \vdots \\
    \vdots \\
   	\phantom{0} \\
   	(Nv+x)\\
	\phantom{0} \\
    \vdots \\
    \phantom{0} 
    \end{array}.
\end{align*}  
Denote by $\bf X$ the right shift on the position space alone  (generalizing the symbol for one qubit flip $X$ gate), which as a matrix in our coordinate representation is the $N\times N$ matrix
\begin{equation} \label{Xdef}
\mathbf{X}= 
  \begin{bmatrix}
    0&0&0&\cdots &1 \\
    1&\ddots &\ddots& \ddots &\vdots \\
    0 & \ddots&\ddots & 0&0\\
    \vdots&\ddots&1&0 & 0 \\
    0&\cdots&0&1 & 0   
  \end{bmatrix}.
\end{equation}
We may check that $\mathbf{X} \ket{x} = \ket{x+1}$, where $\ket{x}$ is as in Eq.~\eqref{poseq}.
 The matrix for propagation $\sigma$ on the joint velocity-position space is  a controlled shift of position $\ket{x}$ by velocity $\ket{v}$,  with  $\ket{v}=\ket{+1}$ affecting the right shift $\mathbf{X}$ and  $\ket{v} = \ket{-1}$ affecting  the left shift ${\mathbf{X}}^{\top}$. Its matrix is
\begin{equation} \label{sigmaeq}
\sigma= 
  \begin{bmatrix}
  \mathbf{X} & \bf{0} \\
    \bf{0}&{\mathbf{X}}^{\top}
  \end{bmatrix}.
\end{equation}
where $\bf{0}$ is the matrix of all zeros.

\begin{figure}[H]
\includegraphics[scale=0.8]{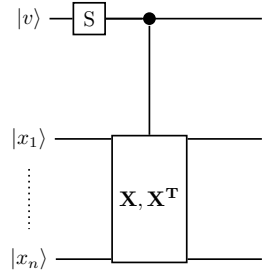}
  \caption{
    A quantum pseudo-circuit for QW. The control $\ket{v}$ selects $\mathbf{X}$ or ${\mathbf{X}}^{\top}$ 
  }
  \label{fig1}
\end{figure}

Figure~\ref{fig1} shows the circuit for this description of QW with  $\hat S$ followed by $\sigma$. 

\subsection{Implementing the  propagation $\sigma$}  \label{sec:sigmaimpl}

The propagation $\sigma$ in Eq.~\eqref{sigmaeq} has a decomposition that we can implement using a single right shift $\mathbf{X}$ from Eq.~\eqref{Xdef}, as we describe next.
Let us first state the definition and a property  of Toeplitz matrices, which we then apply toward our decomposition.  Recall that a  Toeplitz matrix  is a square matrix with each descending diagonal from left to right a constant, i.e., a  $N \times N$ matrix $A$ is Toeplitz if
\begin{equation*} 
A= 
  \begin{bmatrix}
    a_0&a_{-1}&a_{-2}&\cdots &a_{-(N-1)} \\
    a_{1}&a_0 &a_{-1}& \ddots &\vdots \\
    a_{2} &a_{1} & \ddots&\ddots &a_{-2} \\
    \vdots & \ddots&\ddots&\ddots & a_{-1}\\
    a_{N-1}&\cdots &a_{2}&a_{1}&a_0     
  \end{bmatrix}.
\end{equation*}
for some set of numbers $\{a_{-(N-1)},\ldots, a_0, \ldots, a_{N-1}\}$.  So if we number the rows and columns of $A$ by $\{0,\ldots,N-1\}$, then $A_{i,j} = a_{i-j}$ for all $0\le i,j \le N-1$.  Also recall the definition of an {\it exchange} matrix (see Section $1.2.11$ of~\cite{gvl:matcomp}). An  exchange  matrix is a square matrix with all its entries $0$ except those on the anti-diagonal which are all $1$.  Denote the $N\times N$ exchange matrix by $J$,
\begin{align*}
J  &= 
  \begin{bmatrix}
    0&\cdots & 0& 0   &1 \\
    0 &\iddots  &\iddots &1&0 \\
    \vdots&0&\iddots & 0& \vdots\\
    0 & 1&\iddots &\iddots &0\\   
    1 & 0&0&\cdots &0\\   
  \end{bmatrix}.
\end{align*}
The transpose of a Toeplitz matrix can be obtained as follows (Section $4.7$ of~\cite{gvl:matcomp}).
\begin{equation*} 
A^\top = J A J.
\end{equation*}
As the shift matrix $\mathbf{X}$ in Eq.~\eqref{Xdef} is Toeplitz, by the above property, we can express $\sigma$ in Eq.~\eqref{sigmaeq} as
\begin{equation} \label{sigmaeq2}
\sigma= 
  \begin{bmatrix}
{\mathbf{X}} & \bf{0} \\
    \bf{0}& J {\mathbf{X}} J 
    
  \end{bmatrix} =   \begin{bmatrix}
\mathbb{I}_N & \bf{0} \\
    \bf{0}& J
    
  \end{bmatrix}   \begin{bmatrix}
{\mathbf{X}} & \bf{0} \\
    \bf{0}& {\mathbf{X}} 
    
  \end{bmatrix}    \begin{bmatrix}
\mathbb{I}_N & \bf{0} \\
    \bf{0}& J
    
  \end{bmatrix},
\end{equation}
where $J$ is the $N \times N$ exchange matrix and $\mathbb{I}_N$ is the $N \times N$ identity matrix. Let us denote the first (and the last) matrix on the right side product decomposition by $C^v(J)$,
\begin{equation} \label{Cvj}
C^v(J) = \begin{bmatrix}
\mathbb{I}_N & \bf{0} \\
    \bf{0}& J
  \end{bmatrix}.
\end{equation}
Then  $\sigma$ in Eq.~\eqref{sigmaeq2} may be written
\begin{equation} \label{sigmacv}
\sigma = C^v(J) \: (\mathbb{I}_v \otimes \mathbf{X}) \: C^v(J),
\end{equation}
where $\mathbb{I}_v$ is the  identity operator on the velocity space. 

\subsubsection{Specializing to qubits} \label{subsec:nqubits}

Let us assume further  that  the lattice is of size $N=2^n$ for some finite $n$, natural for multi-qubit  systems.  The binary  expansion of  the position  $x$   
\begin{equation} \label{xdec}
x = \sum_{j=0}^{n-1} 2^j x_j,
\end{equation}
  with   $x_j \in \{0,1\}$, $j \in \{0,\ldots,n-1\}$, gives us a bijective map $x \leftrightarrow (x_{n-1} \ldots x_0)$. This   identifies the bases of position space and  of $n$ qubits: $\ket{x} \leftrightarrow \ket{x_{n-1} \ldots x_0}$. 
We also identify  the velocity basis states  with the usual qubit states as   $\ket{+1} \leftrightarrow \ket{0}$, and  $\ket{-1} \leftrightarrow \ket{1}$.   Combined, the velocity-position space is thus identified with  $n+1$ qubits, 
\begin{align*}
\mathcal{H} &\leftrightarrow \mathbb{C}^{n+1} \\ 
\ket{+1,x} &\mapsto  \ket{0, x_{n-1} \ldots x_0}\\
\ket{-1,x} &\mapsto  \ket{1, x_{n-1} \ldots x_0}
\end{align*}
For the rest of this section when we refer to any of the  vector spaces, we interpret that as the corresponding multi-qubit space.

\begin{figure}[H] 
\includegraphics[scale=1]{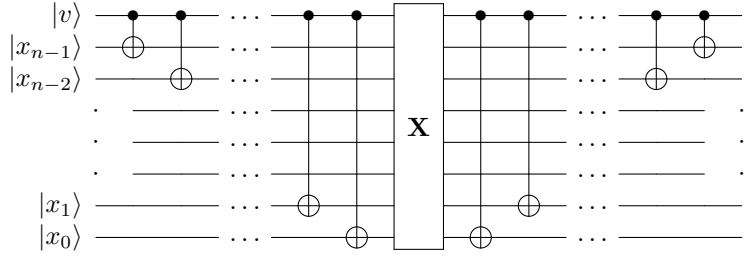}
  \caption{
    $n$-qubit $\sigma$ implementation using a single shift $\mathbf{X}$
  }
  \label{figqinc}
\end{figure}

Let us turn to the operators. We observe that the exchange matrix $J$, in the basis ordering we have,  is simply the qubit-wise  application of $X$ to each position qubit $\ket{x_j}$ (flipping each $\ket{x_j}$: $\ket{0} \rightleftharpoons \ket{1}$),
\begin{equation} \label{Jdecomp}
J = \bigotimes_{j=1}^n X,
\end{equation}
where
\begin{equation*}
X = \begin{bmatrix}
0 & 1 \\
    1& 0
  \end{bmatrix}.
\end{equation*}
Denote by $C^v_j(X)$ the  $\ket{v}$-controlled $X$ on position qubit $\ket{x_j}$, i.e.,  a CNOT gate controlled by $\ket{v}$ with  target $\ket{x_j}$,
\begin{align*}
C^v_j(X)  = 
  \begin{bmatrix}
    \mathbb{I}_2 & \bf{0} \\
	 \bf{0} & X   
  \end{bmatrix},
\end{align*}
where $\mathbb{I}_2$ is the $2\times 2$ identity matrix and  the basis of the  space on which  $C^v_j(X)$ acts is  $\{\ket{v,x_j}\}$. 
By the expression for $J$ in Eq.~\eqref{Jdecomp}, and  because $C^v_j(X)$ and $C^v_j(X)$ commute  for all $i,j$ ,  we can write $C^v(J)$ in eq.~\eqref{Cvj} as
\begin{equation} \label{CVjmq}
C^v(J) = \prod^n_{j=1} C^v_j(X).
\end{equation}
This means that  $\sigma$ in eq.~\eqref{sigmacv} can be implemented as  in Figure~\ref{figqinc}. 
Note that the order in which the CNOT gates are applied on either side of $\mathbf{X}$ in the circuit of Figure~\ref{figqinc} does not matter: they have independent targets and may even be executed in parallel. We have simply chosen one arrangement for the convenience of creating a figure.


By the decomposition  just presented, we have taken a substantial step toward decreasing  the resources needed for implementation of $\sigma$. This is because we use  the  same shift $\mathbf{X}$ for increment and decrement,  instead of  explicitly  implementing  both a controlled increment and decrement circuit. Note that  the  left/right direction control is affected by $C^v(J)$, while the shift $\mathbf{X}$ only works as a right shift (increment). This allows us the flexibility of choosing any shift circuit for $\mathbf{X}$. To reduce the resources most effectively, we consider the  shift implementation that uses the quantum Fourier transform.

\subsection{Quantum Fourier transform  based shift implementation} \label{subsubsec:qftshift}

The quantum Fourier transform (QFT)~\cite{sp:ptfd,nc:qcqi}  diagonalizes the shift $\mathbf{X}$ (see section $4.8$ of~\cite{gvl:matcomp}).  Denote the $N$ dimensional QFT by $\mathcal{F}$,~\footnote{This description of the QFT based shift algorithm was communicated to the author  by   David Meyer.}
\begin{equation*}
\mathcal{F} : \ket{x} \rightarrow \frac{1}{\sqrt{N}} \sum_{k=0}^{N-1} e^{2\pi i x k/N}\ket{k},
\end{equation*}
then
\begin{equation*}
\mathbf{X}=\mathcal{F}^{-1} \Omega \mathcal{F},
\end{equation*}
where $\Omega$ is a diagonal phase multiplication matrix
\begin{align*}
\Omega  = 
  \begin{bmatrix}
    1&0&0&\cdots &0 \\
    0&\omega &\ddots& \ddots &\vdots \\
    0&0&\omega^2&0 & 0 \\
    \vdots & \ddots&\ddots & \ddots&0\\
    0&0&\cdots&0 & \omega^{N-1}    
  \end{bmatrix}.
\end{align*}
This simplifies to a tensor product of $n$ single qubit phase rotations when $N=2^n$,
\begin{align} \label{omegaexp}
\Omega  =   \begin{bmatrix}
    1&0\\
    0 & \omega
  \end{bmatrix} \otimes  \begin{bmatrix}
    1&0\\
    0 & \omega^{2}
  \end{bmatrix} \otimes \begin{bmatrix}
    1&0\\
    0 & \omega^{4}
  \end{bmatrix}   \otimes \cdots \otimes \begin{bmatrix}
    1&0\\
    0 & \omega^{2^{n-1}}
  \end{bmatrix},
\end{align}
where $\omega = e^{2\pi i/2^n}$. 

\begin{figure}[H] 
\includegraphics[scale=0.8]{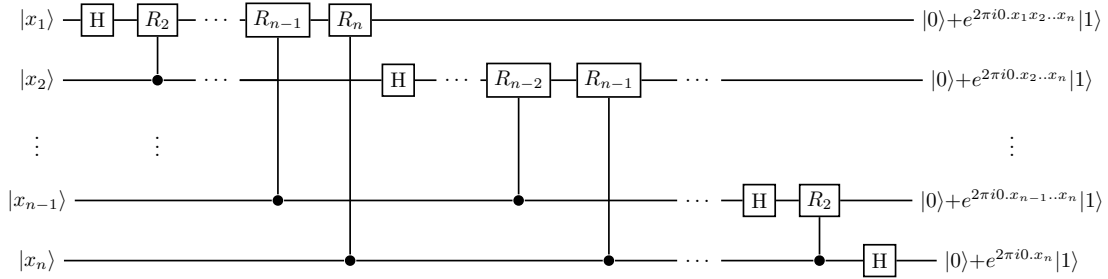}
  \caption{
    Quantum Fourier Transform circuit. 
  }
  \label{fig2}
\end{figure}

The Shor's algorithm~\cite{nc:qcqi} for QFT  as a circuit  is in Figure\ref{fig2}.
where 
\begin{equation} \label{Rkexp}
R_k=\begin{bmatrix}
    1&0\\
    0 & e^{2\pi i/{2^k}}
  \end{bmatrix}.
\end{equation}
Note that, strictly, we need to insert $\lfloor n/2 \rfloor$ swaps (each in turn requiring $3$ CNOT gates) at the end of the  QFT  circuit to get the output ordered correctly. Since we use QFT and the inverse quantum Fourier transform (IQFT) together to achieve the shift $\mathbf{X}$, the swaps can be absorbed in $\Omega$, by reversing the order of phase rotation operations on qubits. To be precise, we can express $\Omega$ in eq.~\eqref{omegaexp} in terms of $R_k$ in eq.~\eqref{Rkexp} above as
\begin{align} \label{omegaexpR}
\Omega  =   R_{n} \otimes  R_{n-1} \otimes R_{n-2}  \otimes \cdots \otimes R_{1}
\end{align}
The swaps in QFT and IQFT can be omitted by replacing $R_k$, $k \in \{1, \ldots, n\}$, in the above expression for $\Omega$ with $R_{n-k+1}$.

Let us  estimate the number of (at most) two-qubit gates needed for this implementation of $\mathbf{X}$. QFT~\cite{nc:qcqi} and its inverse each require  $n(n+1)/2$ gates, excluding swaps at the end, which are not needed by the observation above. $\Omega$ requires $n$ phase rotations.  Thus in total, this implementation of  $\mathbf{X}$ requires  $n^2 +2n$ gates.  The circuit depth is $n$. The entire QW implementation requires an additional $2n$ CNOT gates and a rotation for the scattering. This adds up to a circuit size of $n^2 + 4n +1$ and circuit depth of $3n+1$.

\subsubsection{Comparison of QFT based shift implementation with generalized CNOT  based implementation} \label{subsubsec:cnot}

The  right  shift ${\mathbf{X}}$ can be implemented using generalized CNOT gates, for instance, as in~\cite{dw:eqciqw}. Implementations of generalized  CNOT gates are described in~\cite{nc:qcqi,bbcdmsssw:egqc,mm:ldnqtg}, and the circuit symbol for an $n$-qubit generalized CNOT gate is shown in Figure~\ref{fig3a} below. 
\begin{figure}[H] 
\includegraphics[scale=1]{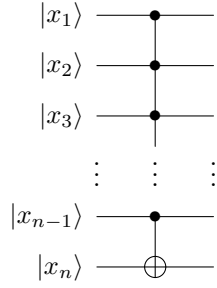}
  \caption{
    $n$-qubit generalized CNOT  gate
  }
  \label{fig3a}
\end{figure}
Shift ${\mathbf{X}}$ implementation using the generalized CNOT gates is shown in Figure~\ref{fig3} below.
\begin{figure}[H] 
\includegraphics[scale=1]{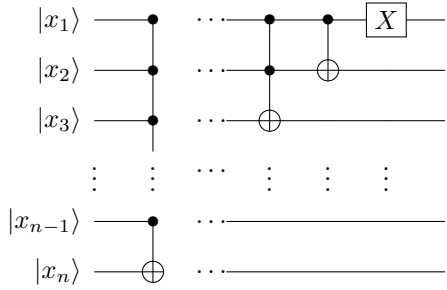}
  \caption{
    Generalized CNOT  gate based $n$-qubit   shift
  }
  \label{fig3}
\end{figure}

When $n=3$  the generalized CNOT gate is also called the CCNOT gate or  the Toffoli gate.  For $n\ge 3$, a generalized CNOT gate  can be implemented~\cite{mm:ldnqtg} by a circuit of size $2n^2 -6n+5$  and depth $8n-20$. The shift circuit above, for $n\ge 3$, without any further circuit optimizations that may be possible, therefore, has size $n(n+1)(2n+1)/3 -3n(n+1) + 5n =  n (2 n^2 -6n + 7)/3$ and depth $4n(n+1)-20n+18 = 2(2n^2-8n+9)$. Both the circuit size and depth in this shift implementation are an order of $n$ in magnitude  larger than the QFT based shift.

An alternative is to use ancilla qubits and CNOT gates as in~\cite{bbcdmsssw:egqc}. For a $n \ge 4$ qubit generalized CNOT gate, this would require $n-3$ ancilla qubits and $4(n-3)$ CCNOT gates, with a circuit depth (in CCNOT gates) of $4(n-3)$. From~\cite{mm:ldnqtg}, each CCNOT gate requires $4$ (at most) $2$-qubit gates.  The generalized CNOT then translates to a a circuit size of   $20(n-3)$ in (at most) $2$-qubit gates  and depth of $16(n-3)$.  The $n-3$  ancilla qubits  can be reused among the generalized CNOT gates of the shift circuit. The shift circuit, for $n\ge 4$, without any further circuit optimizations, would then be of size  $10n^2-50n+67$ and of depth  $2(4n^2-20n+27)$.  This, for large $n$, is a factor of $5$ higher than the QFT based shift in circuit size and an order of $n$ in magnitude larger in depth, even without the cost of ancilla qubits, which effectively double  the number of required qubits in this circuit implementation.

\section{QW evolution with spatial convolution} \label{sec:qftconv}
QFT, in fact,  diagonalizes any unitary circulant matrix (see section $4.8$ of~\cite{gvl:matcomp}), i.e., any convolution matrix. Recall that an $N\times N$  circulant matrix is a special type of Toeplitz matrix of the form
\begin{equation*} 
C= 
  \begin{bmatrix}
    c_0&c_{1}&c_{2}&\cdots &c_{N-1} \\
    c_{N-1}&c_0 &c_{1}& \ddots &\vdots \\
    c_{N-2} &c_{N-1} & \ddots&\ddots &c_{2} \\
    \vdots & \ddots&\ddots&\ddots & c_{1}\\
    c_{1}&\cdots &c_{N-2}&c_{N-1}&c_0     
  \end{bmatrix}.
\end{equation*}
Instead of the dynamics based on shift $\mathbf{X}$, we would like to  accomplish the dynamics of QW based on  unitary circulant matrices. In this scheme, the state at a site after a time step would depend on the state of the walk at the  neighboring  sites before the step through a unitary spatial convolution more general than a shift. Let circulant matrices  $C, C'$ determine the dynamics in that the propagation matrix, in the joint velocity-position~\footnote{We realize that {\it velocity} is no longer an appropriate name for this form of steering.} basis, is
\begin{equation} \label{sigmaeqc}
\sigma= 
  \begin{bmatrix}
  C & \bf{0} \\
    \bf{0}&C'
  \end{bmatrix}.
\end{equation}
The QW evolution in this setting proceeds as before by  scattering  the velocity by a unitary matrix $S$  followed by the  propagation above. The evolution of the QW is:
\begin{equation*}
T = \sigma  \hat S,
\end{equation*}
where $\hat{S}$ is as in eq.~\eqref{hats}. 

 Let
\begin{equation*} 
C=\mathcal{F}^{-1} \Lambda \mathcal{F} = \mathcal{F}^{\dag} \Lambda \mathcal{F}
\end{equation*}
where 
\begin{align*} 
\Lambda  = 
  \begin{bmatrix}
    e^{i \theta_0} &0&0&\cdots &0 \\
    0&e^{i \theta_1} &\ddots& \ddots &\vdots \\
    0&0&e^{i \theta_2}&0 & 0 \\
    \vdots & \ddots&\ddots & \ddots&0\\
    0&0&\cdots&0 & e^{i \theta_{N-1}}    
  \end{bmatrix},
\end{align*}
and  $\theta_j$,  $j \in \{0,\ldots,N-1\}$, are  real.   Similarly we have the decomposition
\begin{equation*} 
C'= \mathcal{F}^{\dag} \Lambda' \mathcal{F},
\end{equation*}
where $\Lambda'$ is a diagonal unitary matrix.

 We can write $\sigma$ in eq.~\eqref{sigmaeqc} as
\begin{equation*} 
\sigma =   
\begin{bmatrix}
\mathcal{F}^{\dag} & \bf{0} \\
    \bf{0}& \mathcal{F}^{\dag}
    
  \end{bmatrix}   \begin{bmatrix}
\Lambda & \bf{0} \\
    \bf{0}& \Lambda'
    
  \end{bmatrix}    \begin{bmatrix}
\mathcal{F} & \bf{0} \\
    \bf{0}& \mathcal{F}
    
  \end{bmatrix}.
\end{equation*} 
We adopt the following notation for $\ket{v}$-controlled switch between matrices $M$ and $M'$ as
\begin{equation*} 
C^v(M, M') = \begin{bmatrix}
M & \bf{0} \\
    \bf{0}& M'
  \end{bmatrix}.
\end{equation*}
To be consistent with the previous section,  we define  $C^v(M) := C^v(\mathbb{I}_N, M)$, and $C^{\bar{v}}(M) := C^v(M,\mathbb{I}_N)$.

 In the above notation, $\sigma$ in Eq.~\eqref{sigmaeqc} may be written
\begin{equation} \label{sigmacvd2}
\sigma = (\mathbb{I}_v \otimes \mathcal{F}^{\dag}) \: C^v(\Lambda, \Lambda') \:  (\mathbb{I}_v \otimes \mathcal{F}) ,
\end{equation}
where $\mathbb{I}_v$ is the  identity operator on the velocity space. 
Insofar as  $\Lambda, \Lambda'$ are  efficiently implementable, and  controllable by $\ket{v}$ to switch between $\Lambda$ and $\Lambda'$, we  can efficiently  implement a QW with its evolution based on $C, C'$ using the QFT. Generally, if we can implement both $C^v(\Lambda, \mathbb{I}_N) = C^{\bar{v}}(\Lambda)$  and $C^v(\mathbb{I}_N, \Lambda') = C^v(\Lambda')$, we can obtain $C^v(\Lambda, \Lambda')$ as
\begin{equation*}
 C^v(\Lambda, \Lambda') = C^v(\Lambda, \mathbb{I}_N) \: C^v(\mathbb{I}_N, \Lambda') = C^{\bar{v}}(\Lambda) \: C^v(\Lambda') =  C^v(\Lambda')  \: C^{\bar{v}}(\Lambda).
\end{equation*} 
The last equality is there as  the controlled  operations may be applied in either order.
Let us re-examine the basic QW as an  illustrative example of the setup of this section in a multi-qubit system  and then consider the general case.

\subsection{Basic QW in a multi-qubit system revisited} \label{subsec:convq}
For the basic QW,  $C=\mathbf{X}$ and $C'=\mathbf{X}^\top$, with  $\Lambda = \Omega$ from eq.~\eqref{omegaexpR} and $\Lambda' = \overline{\Omega}$, the element-wise complex conjugate of $\Omega$, 
\begin{align*} 
\overline{\Omega}  =   \overline{R}_{n} \otimes  \overline{R}_{n-1} \otimes \overline{R}_{n-2}  \otimes \cdots \otimes \overline{R}_{1},
\end{align*}
where $R_j$ are as in eq.~\eqref{Rkexp},
\begin{equation*} 
R_j=\begin{bmatrix}
    1&0\\
    0 & e^{2\pi i/{2^j}}
  \end{bmatrix},
\end{equation*}
and
\begin{equation*} 
\overline{R}_j=\begin{bmatrix}
    1&0\\
    0 & e^{-2\pi i/{2^j}}
  \end{bmatrix}.
\end{equation*}
We start with the standard  controlled rotations (with $\ket{v}$ as the control),
\begin{align*}
C^v_j(\overline{R}_j)  = 
  \begin{bmatrix}
    \mathbb{I}_2 & \bf{0} \\
	 \bf{0} & \overline{R}_j   
  \end{bmatrix},
\end{align*}
and
\begin{align*}
C^{\bar{v}}_j(R_j)  = 
  \begin{bmatrix}
    R_j &  \bf{0} \\
	 \bf{0} &  \mathbb{I}_2  
  \end{bmatrix}.
\end{align*}
Note that 
\begin{align} \label{vtovbar}
C^{\bar{v}}_j(R_j)  = (X \otimes \mathbb{I}_2) \: C^v_j(R_j) \:  (X \otimes \mathbb{I}_2)
\end{align}
Using these, we can construct 
\begin{equation*} \label{CVjmq}
C^{\bar{v}}(\Omega) = \prod^n_{j=1} C^{\bar{v}}_j(R_j),
\end{equation*}
and 
\begin{equation*} \label{CVjmq}
C^v(\overline{\Omega}) = \prod^n_{j=1} C^v_j(\overline{R}_j),
\end{equation*}
and thus
\begin{equation*}
 C^v(\Omega, \overline{\Omega}) =  C^{\bar{v}}(\Omega) \: C^v(\overline{\Omega})
\end{equation*}
Substituting  in ~\eqref{sigmacvd2}, we get the propagation $\sigma$ in the setup of this section,
\begin{equation*} \label{sigmaomega}
\sigma = (\mathbb{I}_v \otimes \mathcal{F}^{\dag}) \: C^v(\Omega,\overline{\Omega}) \:  (\mathbb{I}_v \otimes \mathcal{F}).
\end{equation*}
This form of $\sigma$ gives us another algorithm for  the basic QW.

\subsection{Multi-qubit system:  the general case} \label{subsec:convqgc}

Let us turn to the general unitary diagonal matrices $\Lambda, \Lambda'$ of eq.~\eqref{sigmacvd2} for the multi-qubit system.  In~\cite{wgma:eqcduwa}, a method is described to  efficiently implement multi-qubit diagonal unitary matrices, using CNOT gates and single qubit rotations of the form
\begin{equation} \label{rotmat}
R_{\theta}=e^{-i \frac{\theta}{2} Z} = \begin{bmatrix}
e^{-i \frac{\theta}{2}} & 0 \\
    0& e^{i \frac{\theta}{2}}
  \end{bmatrix},
\end{equation}
where 
\begin{equation*}
Z=\begin{bmatrix}
1 & 0 \\
    0& -1
  \end{bmatrix}.
\end{equation*}
The approximation in~\cite{wgma:eqcduwa} uses partial Walsh-Fourier series. The circuit size depends on  the approximation error $\epsilon$  as $O(1/\epsilon)$, and is   independent of $n$ for large enough $n$.
That implementation  can be utilized with a simple modification to   implement $C^v(\Lambda)$ efficiently.  
Given  an efficient  circuit prescribed in~\cite{wgma:eqcduwa} for the diagonal unitary matrix $\Lambda$, suppose that some  qubit in the circuit has a rotation of the form $R_{\theta}$ from eq.~\eqref{rotmat} applied to it.  We replace this rotation with  $C^v(R_{\theta})$, the  $\ket{v}$-controlled $R_{\theta}$ applied to that qubit,
\begin{align*}
C^v(R_{\theta}) = 
  \begin{bmatrix}
    \mathbb{I}_2 & \bf{0} \\
	 \bf{0} & R_{\theta}   
  \end{bmatrix},
\end{align*}  
Substituting $\ket{v}$-controlled versions for all such rotations on all qubits would give us  $C^v(\Lambda)$. In a similar manner, we can obtain  $C^{\bar{v}}(\Lambda)$ by substituting $C^{\bar{v}}(R_{\theta})$, which can be constructed as in eq.~\eqref{vtovbar} if necessary, for  the rotations $R_{\theta}$ that occur in the circuit for $\Lambda$. 
With  $C^{\bar{v}}(\Lambda)$ and $C^{v}(\Lambda')$ thus constructed, we obtain
\begin{equation*}
 C^v(\Lambda, \Lambda') = C^{\bar{v}}(\Lambda) \: C^v(\Lambda') =  C^v(\Lambda')  \: C^{\bar{v}}(\Lambda).
\end{equation*} 
We  use this efficient implementation of  $C^v(\Lambda, \Lambda')$ and the $2^n$ dimensional QFT implementation  to  obtain an efficient implementation of  $\sigma$ from  eq.~\eqref{sigmacvd2}.   
An efficient  QW algorithm for the given $\Lambda, \Lambda'$ follows easily.


%



\section{Simulating  QW with QFT based shift on quantum computers}\label{sec:ibmsim}

We test the QW algorithm just described with the form of propagation from Section~\ref{sec:sigmaimpl} and QFT based shift   on  three IBM quantum computers~\cite{ibm:qexp}: IBMQX$2$,  IBMQX\_London and IBM\_$16$\_Melbourne, which are $5$, $5$ and $14$   qubit machines respectively. 
The reader may find  the machine gate and coupling maps  in Appendix~\ref{appdx:ibmqcl} Section~\ref{subsec:gatecoupmap}. IBM Qiskit~\cite{qiskiturl} provides the API to access the computers and develop  the python based code, transpile it, and run it.

The lattice sizes, walk step pairs  in our simulations are  $(4,1)$, $(4,2)$,  and $(8,1)$ corresponding to $n=2,3$ qubits. For all the simulations, we use the same scattering matrix
\begin{equation*}
S=\frac{1}{\sqrt{2}}\begin{bmatrix}
1&i\\
i&1
\end{bmatrix}.
\end{equation*}  
After initializing the walk joint velocity-position state, we run the algorithm  for the given number of evolutionary steps. Then we   measure the  final state.  

 For each experiment, i.e., a set of lattice size,  initial state, and steps of evolution, we execute  the algorithm $1024$ times (called ``shots") to generate a distribution on measured states.~\footnote{For the terminology in this paper  related to the IBM quantum computers,  the reader might wish to look up IBM quantum computing website~\cite{ibm:qexp}.}   The  circuit is transpiled  (compiled to quantum gates)  using optimization level $3$ of the transpiler prior to executing each experimental run. At this level, the transpiler takes in to account the noise properties and the connectivity of the device.
 
 We repeat these runs, for each experiment and each machine,  numerous times.  We show the result from the run with the smallest $\ell^1$ distance from the ideal distribution.  Note that the $\ell^1$ distance between two probability distributions $P$ and $Q$ over a finite, discrete variable $i \in \mathcal{I}$ is
\begin{equation*}
\ell^1(P,Q) = \frac{1}{2} \sum_{i \in \mathcal{I}} \abs{P(i) - Q(i)}.
\end{equation*}
It takes values in $[0,1]$ range.  For each experiment, we group together the plots of  the distributions for all the  machines, and also the ideal distribution that the QW state would have starting from the same state and after the same number of evolutionary steps. A table after each group of plots records  the circuit size, the circuit depth, and the $\ell^1$ distance between the ideal distribution and the ones shown. We point out that the circuits that give the distributions with the smallest $\ell^1$ distance from the ideal do not always have the smallest transpiled size. We record the smallest  circuit sizes in Appendix~\ref{appdx:ibmqcl} Section~\ref{subsec:sizetrans}.

The first  set of plots are for $N=4$ sites ($n=2$), and  $1$ step of walk evolution.  Recall that the joint velocity-position state is $\ket{v, x}$.     As a reminder, the  joint velocity-position state is $\ket{v, x}$.     For example, the state $\ket{011}$ has $\ket{v}= \ket{0}$, and $\ket{x}= \ket{11}$.   The initial walk state for this set of experiments is chosen to be  $\ket{v, x} = \ket{010}$. 

The ideal plot shows that  the walk has moved  both  right and the left by $1$ lattice point, having been scattered in both a left and  a right moving component. The measured distributions show good agreement with the expected distribution for  IBMQX$2$ and IBMQX\_London, but there are other states sampled as well .  The distribution from IBM\_$16$\_Melbourne has stronger deviation from the ideal. ~\footnote{We can  speculate about the causes and sources of the deviations from the ideal.  The circuit size and depth, gate noise, leakage,  actual sequence of operations and  decoherence, and other factors, are likely.} 
 The circuit layouts for each case are in Appendix~\ref{appdx:ibmqcl} Section~\ref{subsec:gatelayoutq}.

\begin{figure}[H]
  \centering
  \begin{tabular}{@{}c@{}}
\minipage{0.5\textwidth} 
\center 
     \caption*{ideal} 
  \includegraphics[width=0.85\linewidth]{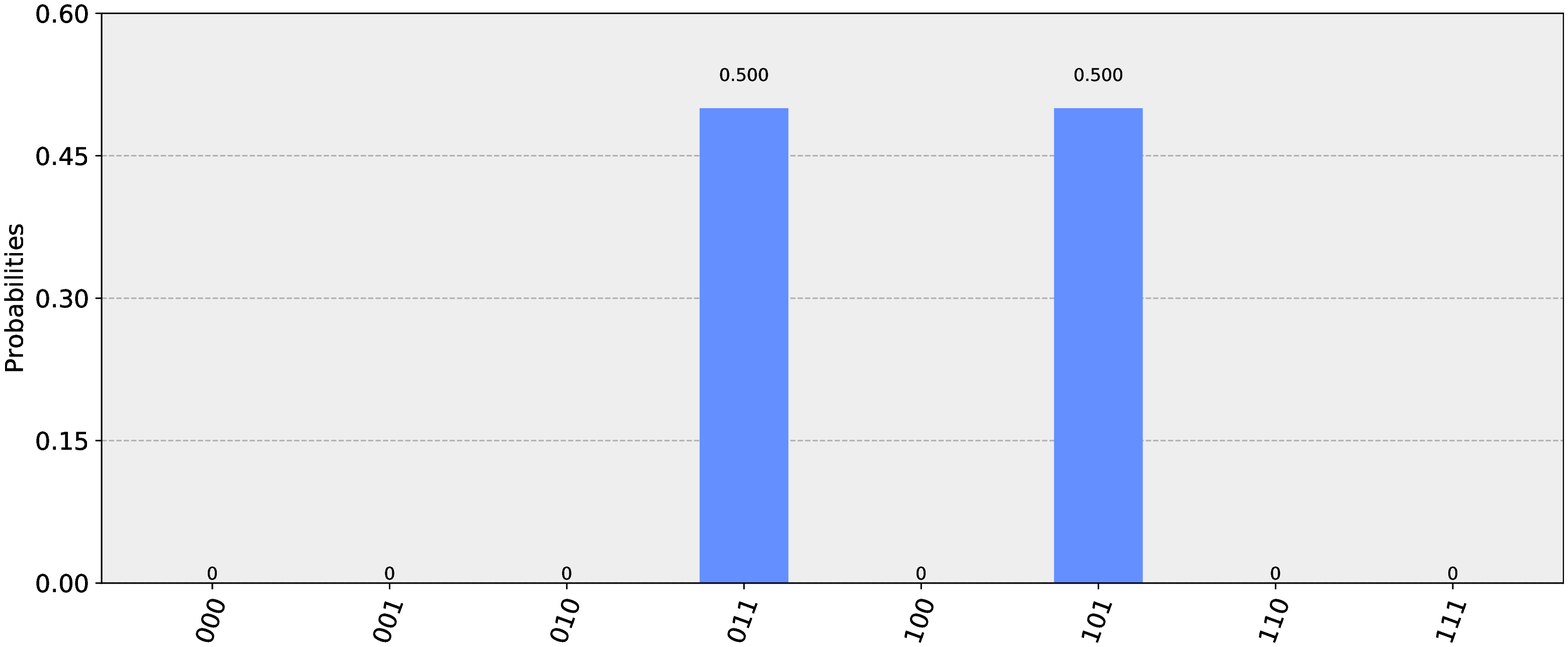} 
\endminipage\hfill
\minipage{0.5\textwidth}
\center
     \caption*{ibmqx2}
  \includegraphics[width=0.85\linewidth]{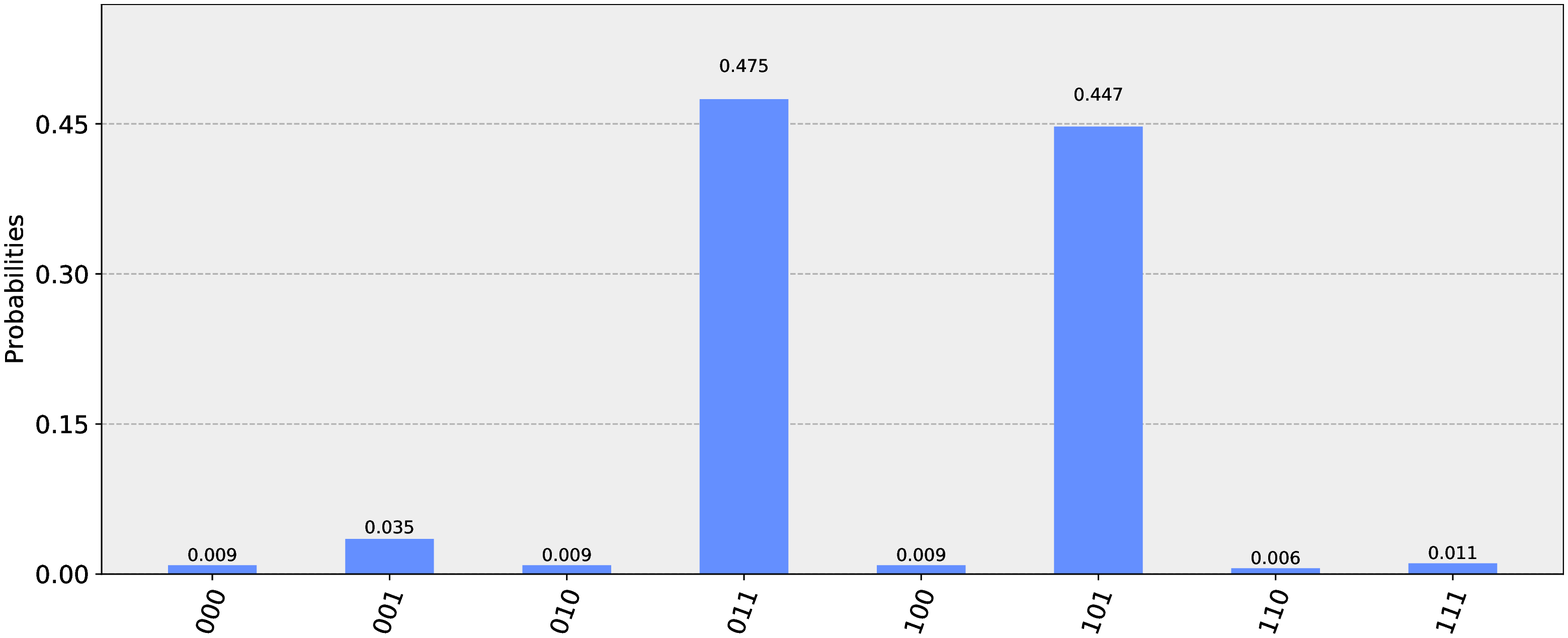} 
 \endminipage
  \end{tabular}


    \begin{tabular}{@{}c@{}}
  \minipage{0.5\textwidth}
\center
     \caption*{ibmq\_london}
  \includegraphics[width=0.85\linewidth]{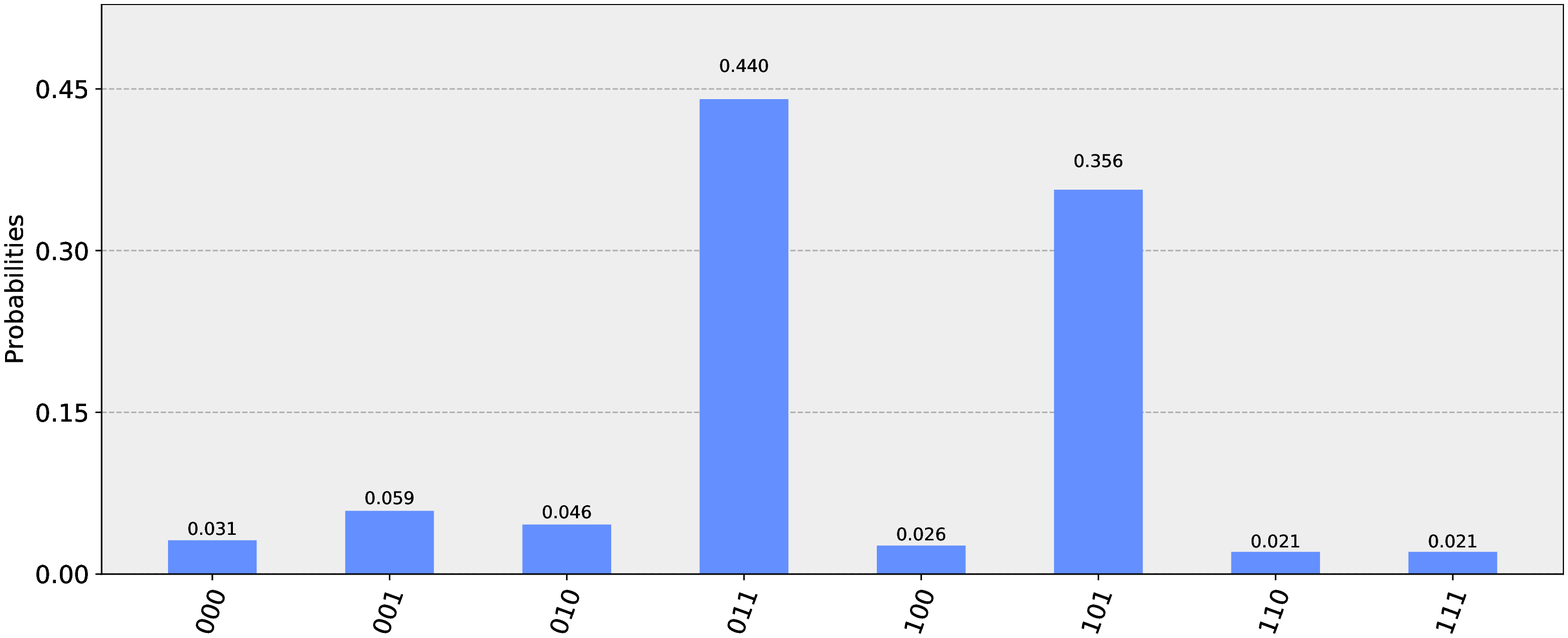} 
 \endminipage
\minipage{0.5\textwidth}
\center 
      \caption*{ibmq\_$16$\_melbourne} 
 \includegraphics[width=0.85\linewidth]{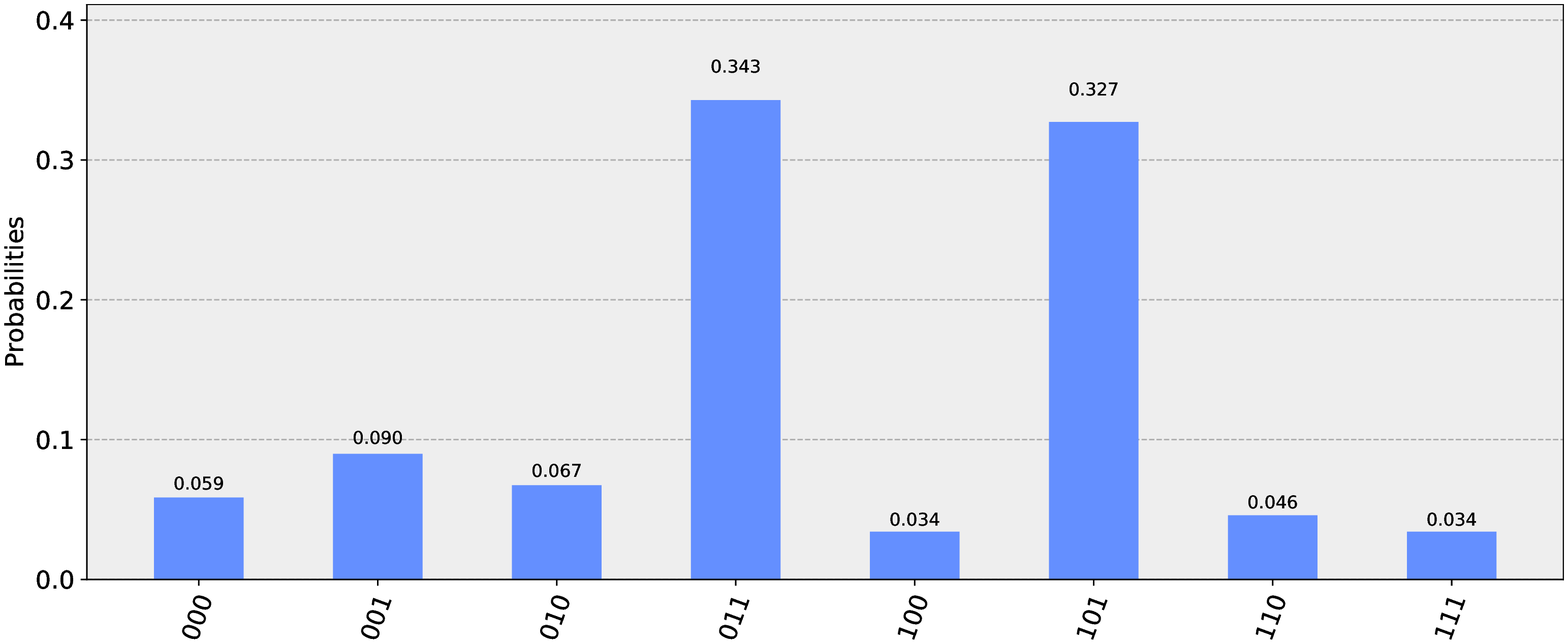} 
\endminipage
  \end{tabular}
  \caption{QFT based shift, $4$ sites, $1$ step}
  \label{fig:q21}
\end{figure}


The table below summarizes the circuit size and  depth,  and the $\ell^1$ distance of the measured distribution from the ideal distribution. It shows that different architectures transpile to different sizes (gate counts) and depths. Comparison with the gate coupling maps in Appendix~\ref{appdx:ibmqcl} confirms  the highest connectivity gate map, which is that of IBMQX2,  corresponds to the lowest circuit size and depth, and the least deviation from the ideal distribution.

\begin{table}   [H]
\begin{center}
\begin{tabu}to\linewidth{|[1.5pt]c|c|c|[1.5pt]}  %
\tabucline[1.5pt]-
& (size , depth) & $\ell^1$ distance from ideal\\
\tabucline[1.5pt]-
ibmqx2 &($11$, $7$)& $0.0781$ \\
\hline 
ibmq\_london &($26$, $18$)& $0.2031$  \\
\hline 
ibmq\_$16$\_melbourne &($33$, $21$)& $0.3301$ \\ 
\tabucline[1.5pt]-
\end{tabu}
\end{center}
\caption{QFT based shift, $4$ sites, $1$ step}
 \label{table:q21}  
\end{table} 

The next  set of plots and the table are for $N=4$ sites ($n=2$), for $2$ steps of walk evolution. Initial state of the walk is $\ket{v, x} = \ket{010}$.
\begin{figure}[H]
  \centering
  \begin{tabular}{@{}c@{}}
\minipage{0.5\textwidth} 
\center 
     \caption*{ideal} 
  \includegraphics[width=0.85\linewidth]{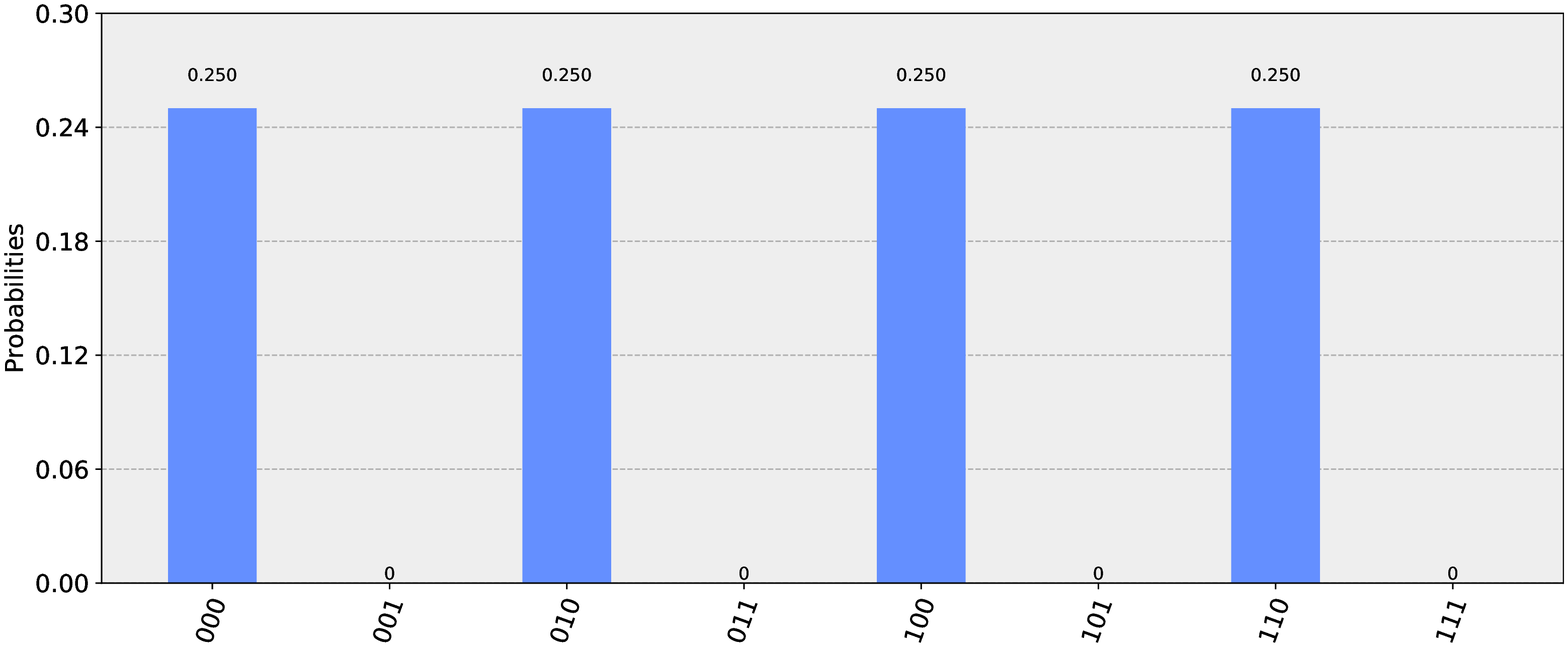} 
\endminipage\hfill
\minipage{0.5\textwidth}
\center
     \caption*{ibmqx2}
  \includegraphics[width=0.85\linewidth]{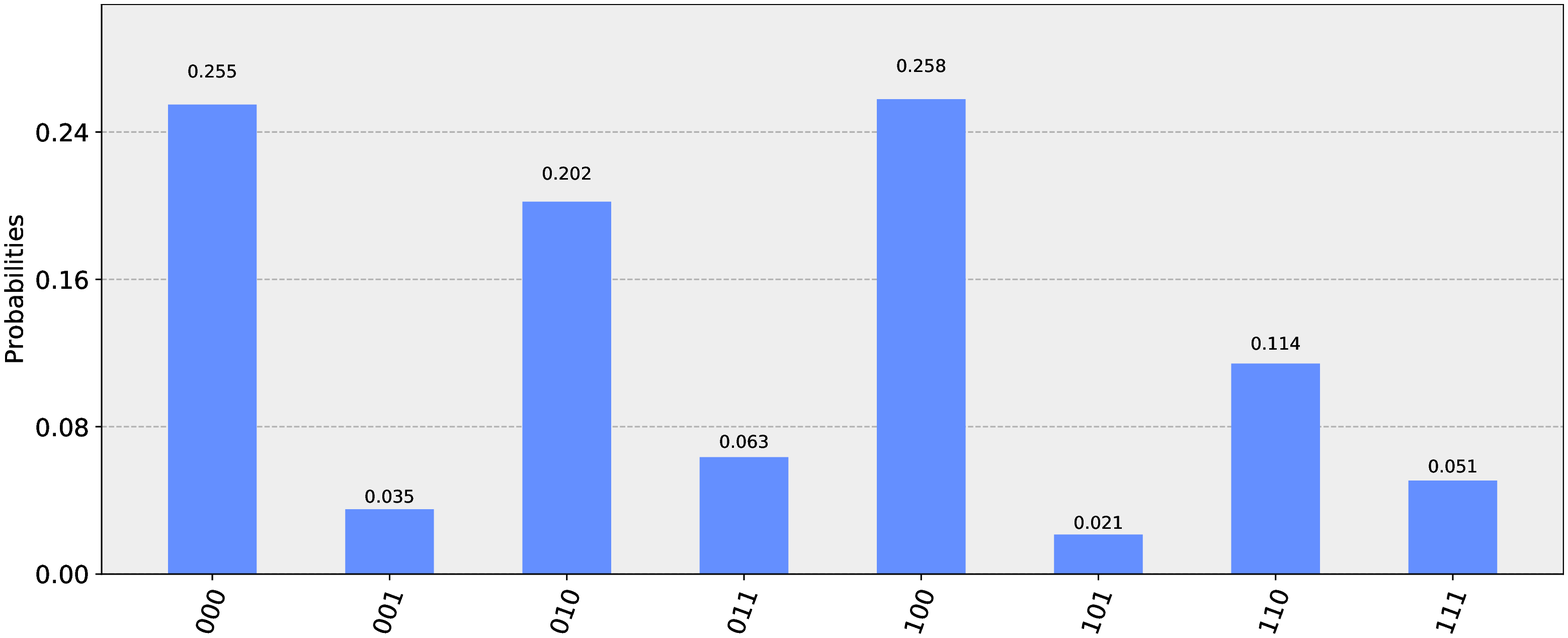} 
 \endminipage
  \end{tabular}
\begin{tabular}{@{}c@{}}
\minipage{0.5\textwidth}
\center
     \caption*{ibmq\_london}
  \includegraphics[width=0.85\linewidth]{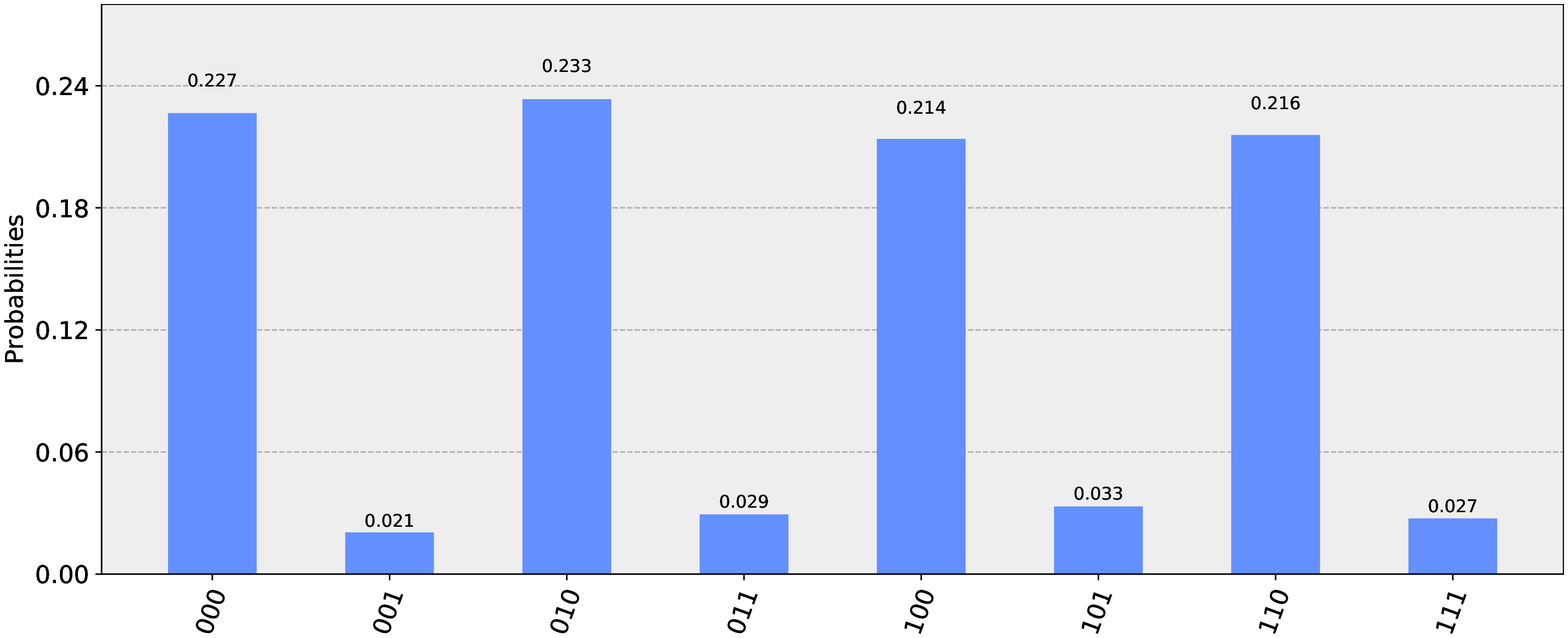}  
 \endminipage
\minipage{0.5\textwidth}
\center 
     \caption*{ibmq\_$16$\_melbourne} 
  \includegraphics[width=0.85\linewidth]{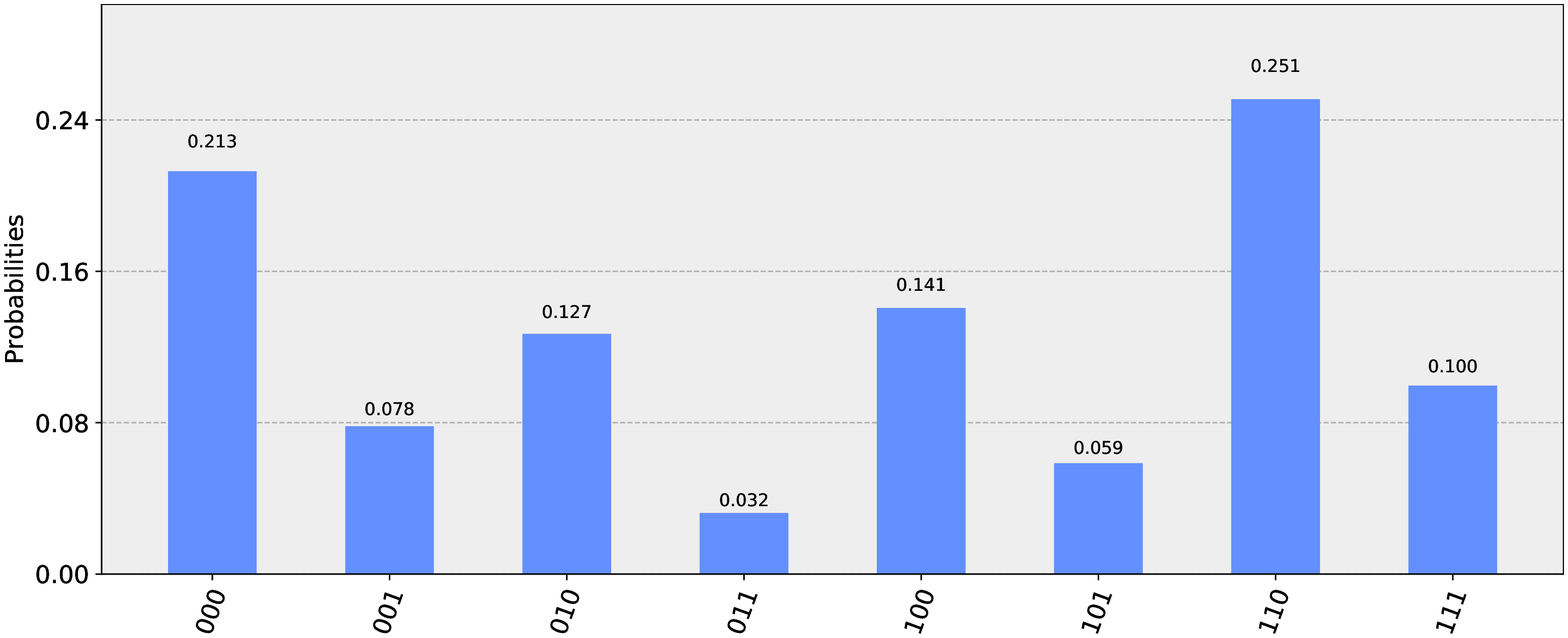}  
\endminipage
  \end{tabular}
  \caption{QFT based shift, $4$ sites, $2$ steps}
  \label{fig:q22}
\end{figure}

The $\ell^1$ distances in the table  mimic the trend in the last section with generalized CNOT based shift. Interestingly, the values are   lower  for IBMQX\_London and IBM\_$16$\_Melbourne than those in the  $1$ step case.

\begin{table}   [H]
\begin{center}
\begin{tabu}to\linewidth{|[1.5pt]c|c|c|[1.5pt]}  %
\tabucline[1.5pt]-
& (size , depth) & $\ell^1$ distance from ideal \\
\tabucline[1.5pt]-
ibmqx2 &($18$, $11$)& $0.1836$\\
\hline 
ibmq\_london &($55$, $35$)& $0.1104$ \\
\hline 
ibmq\_$16$\_melbourne &($57$, $36$)& $0.2695$ \\
\tabucline[1.5pt]-
\end{tabu}
\end{center}
\caption{QFT based shift, $4$ sites, $2$ steps}
 \label{table:q22}  
\end{table} 

This last set of plots are for $N=8$ sites ($n=3$), and  $1$ step of walk evolution. Initial state of the walk is $\ket{v, x} = \ket{0010}$. While the distributions for IBMQX$2$ and IBMQX\_London show reasonable  agreement with the ideal, the distribution from IBM\_$16$\_Melbourne is no longer recognizable as being a result of the quantum walk with discernible peaks at the correct states.

\begin{figure}[H]
  \centering
  \begin{tabular}{@{}c@{}}
\minipage{0.5\textwidth}
\center 
     \caption*{ideal} 
  \includegraphics[width=0.85\linewidth]{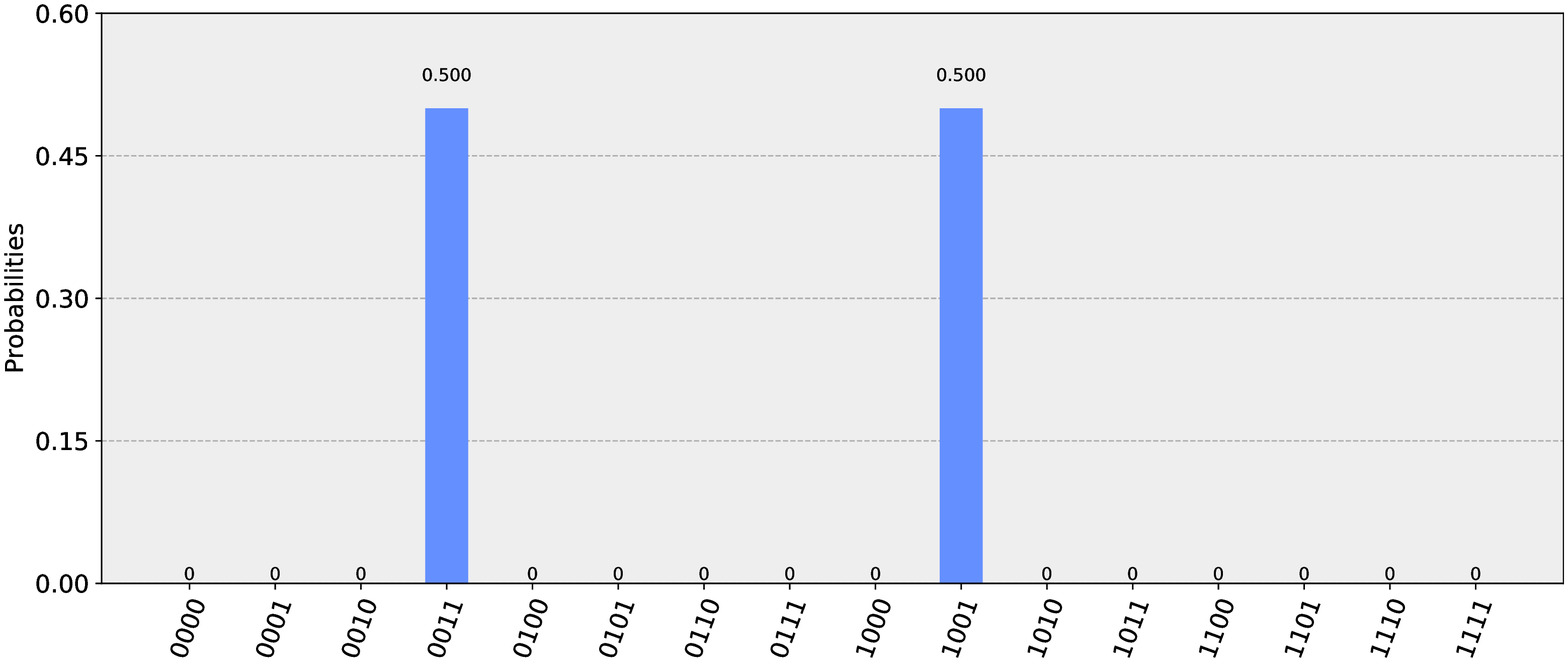}  
\endminipage\hfill
\minipage{0.5\textwidth}
\center
     \caption*{ibmqx2}
  \includegraphics[width=0.85\linewidth]{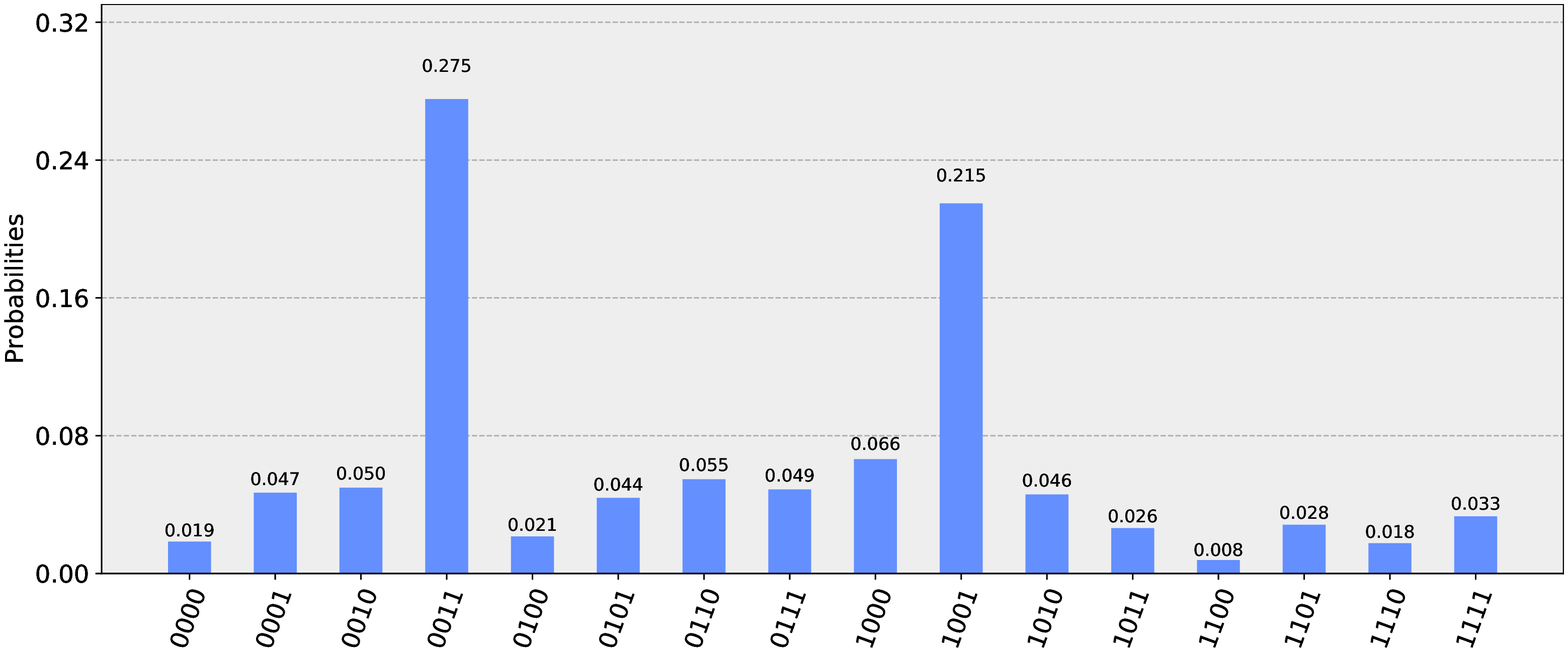} 
 \endminipage
  \end{tabular}
  \begin{tabular}{@{}c@{}}
\minipage{0.5\textwidth}
\center
     \caption*{ibmq\_london}
  \includegraphics[width=0.85\linewidth]{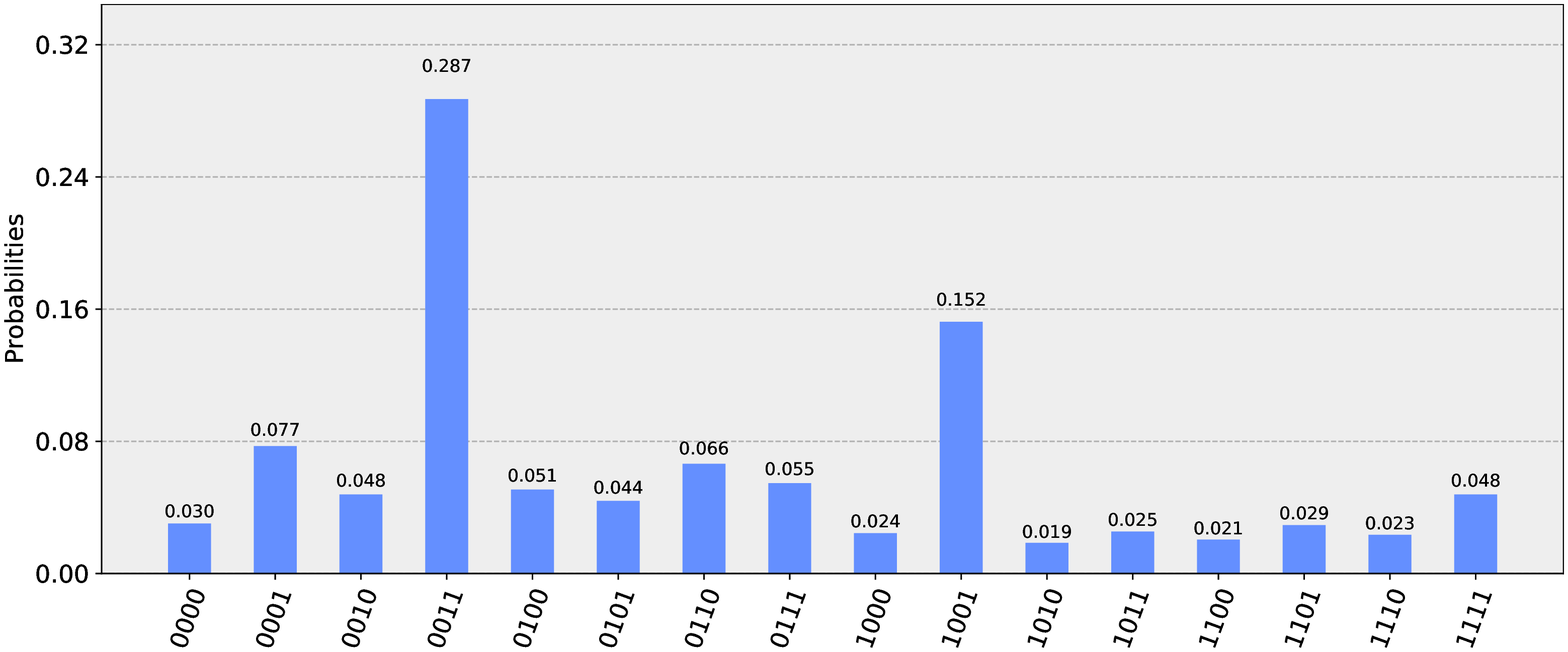} 
 \endminipage
\minipage{0.5\textwidth} 
\center 
     \caption*{ibmq\_$16$\_melbourne} 
  \includegraphics[width=0.85\linewidth]{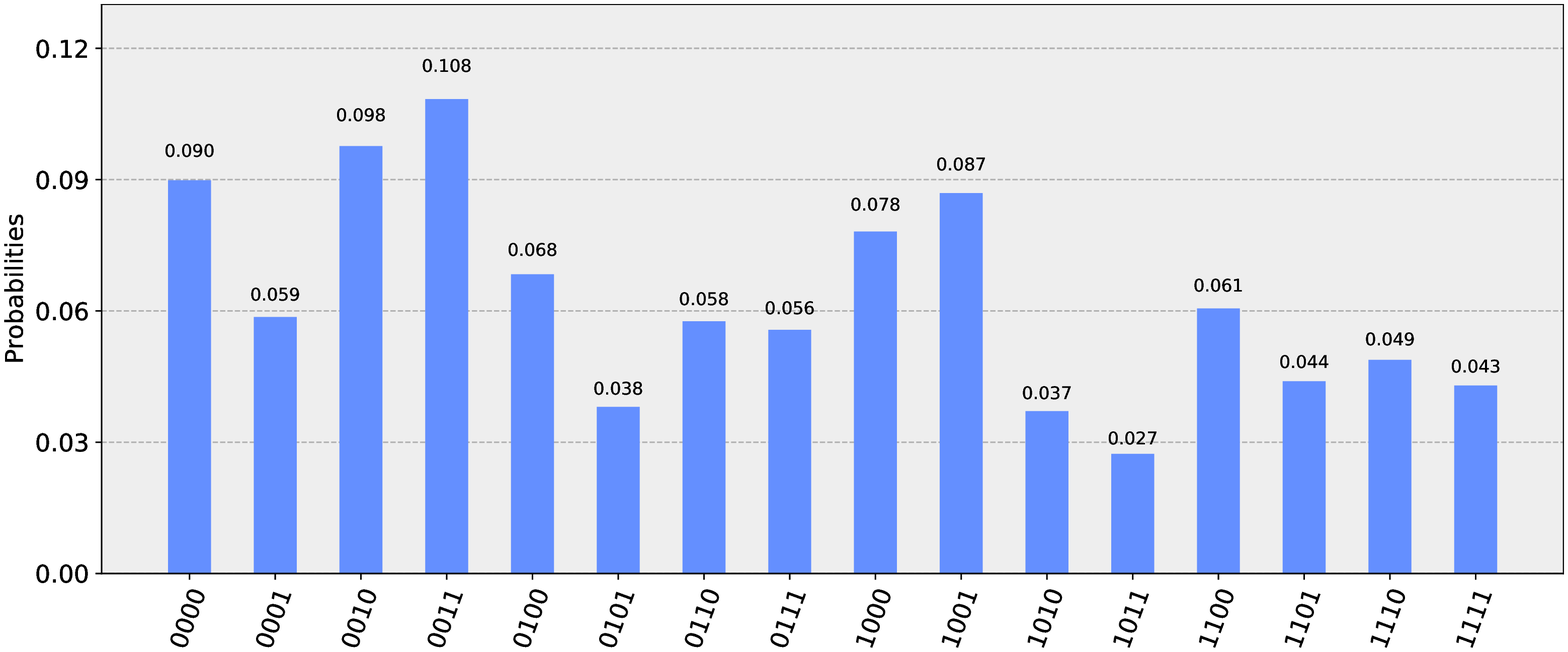} 
\endminipage
  \end{tabular}
  \caption{QFT based shift, $8$ sites, $1$ step}
  \label{fig:q31}
\end{figure}

Overall, as apparent in the table below, the circuits are larger, respectively, than those encountered so far. The  deviations from the  ideal are more pronounced as well. IBM\_$16$\_Melbourne distribution is very distant from the ideal.

\begin{table}   [H]
\begin{center}
\begin{tabu}to\linewidth{|[1.5pt]c|c|c|[1.5pt]}  %
\tabucline[1.5pt]-
& (size , depth) & $\ell^1$ distance from ideal \\
\tabucline[1.5pt]-
ibmqx2 &($62$, $35$)& $0.5098$ \\
\hline 
ibmq\_london &($59$, $38$)& $0.5605$ \\
\hline 
ibmq\_$16$\_melbourne &($79$, $50$)& $0.8047$ \\ 
\tabucline[1.5pt]-
\end{tabu}
\end{center}
\caption{QFT based shift, $8$ sites, $1$ step}
 \label{table:q31}  
\end{table} 

\section{Conclusion}\label{sec:conc}

In this paper we have developed a QW algorithm that uses fewer resources by simplifying the structure  of the controlled shift used to implement the propagation part of a QW. A single shift (increment) circuit suffices as opposed to both an increment and a decrement circuit that would  usually be needed. The implementation allows any shift circuit, so that an optimized shift circuit dependent on the particular machine architecture may be substituted.  In the NISQ regime where noise and decoherence   adversely affect  performance, this serves as an advantage. In comparison to a generalized CNOT based shift, a QFT based shift is generally an order of magnitude  in $n$, the number or qubits, smaller and less deep  and does not use ancilla qubits.

We examine and simulate the QW based on a QFT based shift.   The algorithm  was run on three IBM quantum computers: IBMQX$2$,  IBMQX\_London and IBM\_$16$\_Melbourne, with $5$, $5$ and $14$   qubit architectures respectively. We were able to simulate QW over  small lattice sizes $N=4, 8$ ($n=2, 3$) and for $1$ and $2$  QW evolution steps, with reasonable results. We found that the higher connectivity architectures, like IBMQX$2$ and  IBMQX\_London, generally  perform better,  though   the transpiler optimizations and  the machine architecture and noise properties have a strong influence in determining the  optimal circuit size and depth. We executed each run for $1024$ shots, noting that  a higher a number of shots would result in measurement distributions that are more accurate with respect to the  machine behavior, especially for higher $n$.

We have implemented the simple QW based on a shift in this paper, but the method generalizes to circulant matrices (convolutions). The implementation of the latter would require care in implementing the controlled diagonal unitary matrices needed in  propagation, for which we have indicated a method based on~\cite{wgma:eqcduwa}.  Future work could be directed toward QWs in higher dimensional lattices, aiming to characterize  the QWs with spatial convolution that could have potential applications in quantum search, and in simulating  models of  physics. It could also seek to understand  how the transpiler optimizations could be directed to improve the QW performance on specific architectures. In a similar vein, and more generally, quantum algorithms and simulation models could be adapted by mathematical structure to better meet the  resource constraints of evolving  quantum computers.

\section*{Acknowledgments} 
The author would like to thank  David Meyer  for the ideas and discussions that motivated,  and are partly  represented  in, this paper.

\newpage 

\appendix
\section{Coupling maps, layouts and samllest transpiled circuit sizes} \label{appdx:ibmqcl}

\subsection{Coupling maps} \label{subsec:gatecoupmap}

These are the IBM quantum computer coupling maps for the machines in this paper. The  directional arrows show source-target pair for controlled operations. Arrows in both directions mean both qubits can serve as  the control and target.

\begin{figure}[H]
\center
  \includegraphics[width=0.3\linewidth]{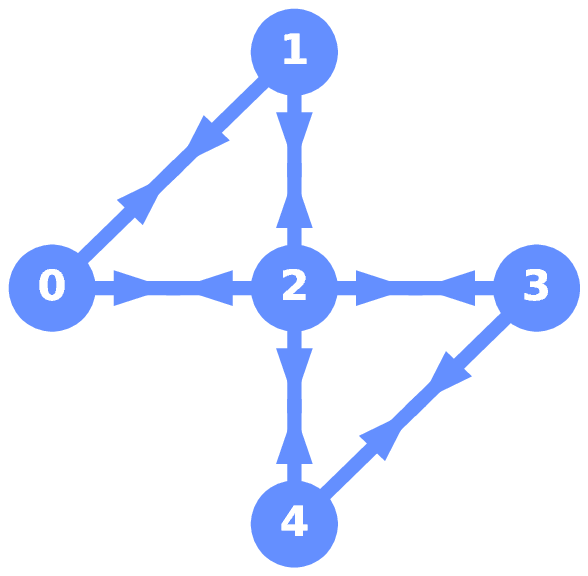} 
     \caption{ibmqx2 coupling map}  
  \label{fig:x2gcmap}
\end{figure}

\begin{figure}[H]
\center
  \includegraphics[width=0.3\linewidth]{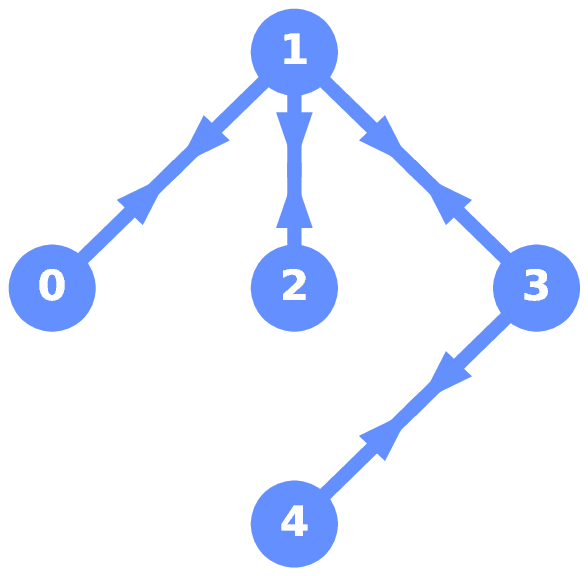} 
     \caption{ibmq\_london  coupling map}  
  \label{fig:x2gcmap}
\end{figure}

\begin{figure}[H]
\center
  \includegraphics[width=0.5\linewidth]{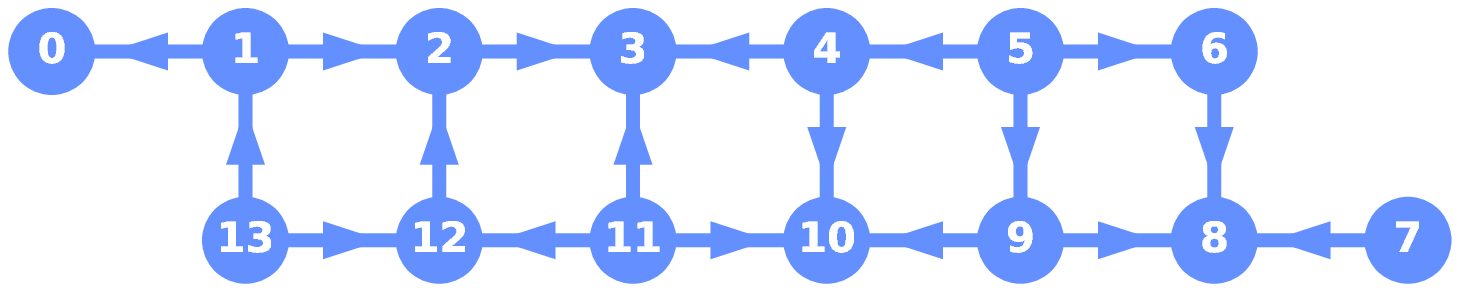} 
     \caption{ibmq\_16\_melbourne coupling map}  
  \label{fig:x2gcmap}
\end{figure}

\newpage

\subsection{Layouts for QFT based shift}  \label{subsec:gatelayoutq}

Here we show the layout maps for each of the experiments with QFT based shift. The qubits and couplings in black are involved in the respective circuit.

\begin{figure}[H]
  \centering
  \begin{tabular}{@{}c@{}}
\minipage{0.3\textwidth} 
\center 
  \includegraphics[width=0.6\linewidth]{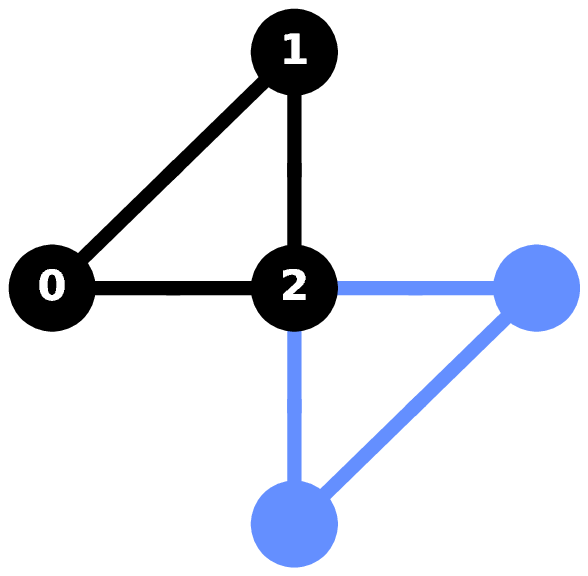} 
     \caption*{ibmqx2} 
\endminipage\hfill
\minipage{0.3\textwidth}
\center
  \includegraphics[width=0.6\linewidth]{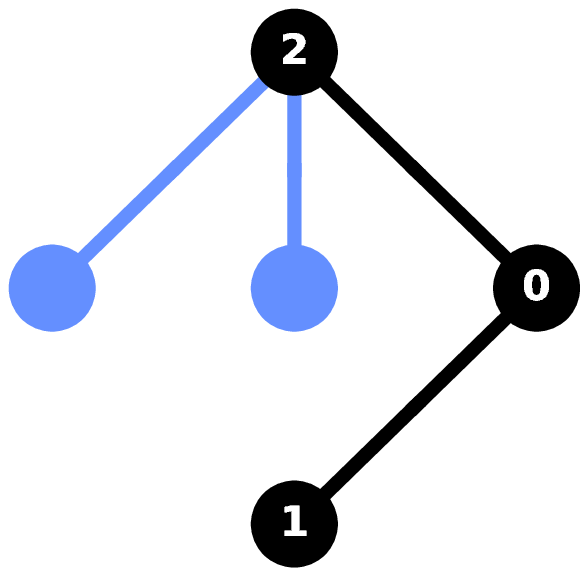} 
     \caption*{ibmq\_london}
 \endminipage\hfill
\minipage{0.3\textwidth}
\center
  \includegraphics[width=1\linewidth]{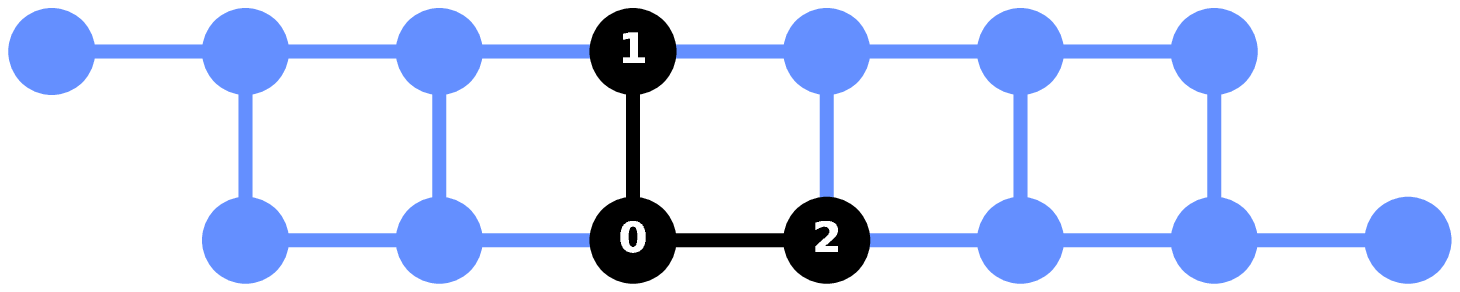} 
     \caption*{ibmq\_16\_melbourne}
 \endminipage   
  \end{tabular}
  \caption{QFT based shift, $4$ sites, $1$ step.}
  \label{fig:lq21}
\end{figure}

\begin{figure}[H]
  \centering
  \begin{tabular}{@{}c@{}}
\minipage{0.3\textwidth} 
\center 
  \includegraphics[width=0.6\linewidth]{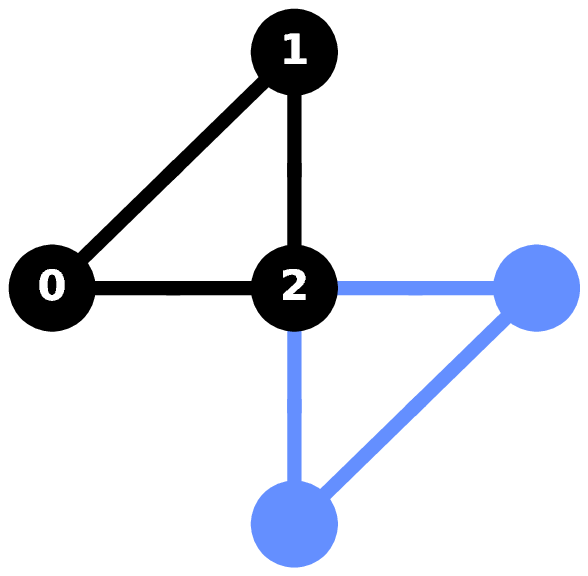} 
     \caption*{ibmqx2} 
\endminipage\hfill
\minipage{0.3\textwidth}
\center
  \includegraphics[width=0.6\linewidth]{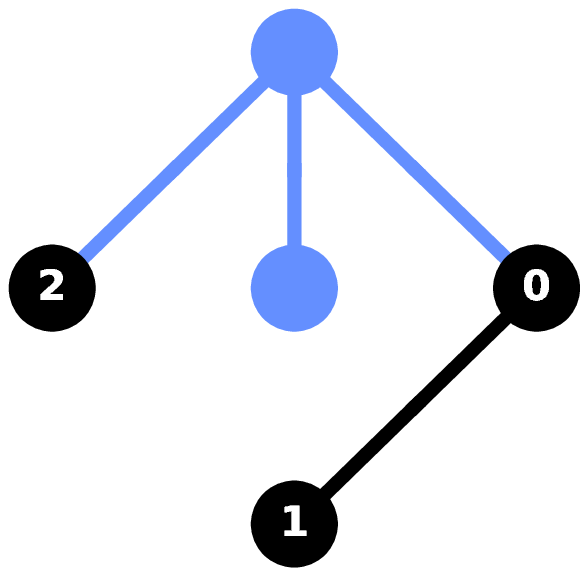} 
     \caption*{ibmq\_london}
 \endminipage\hfill
\minipage{0.3\textwidth}
\center
  \includegraphics[width=1\linewidth]{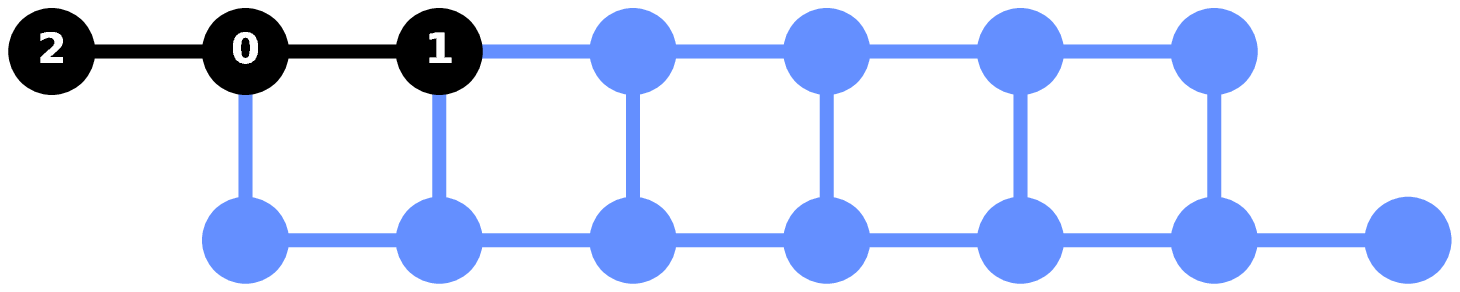} 
     \caption*{ibmq\_16\_melbourne}
 \endminipage   
  \end{tabular}
  \caption{QFT based shift, $4$ sites, $2$ steps.}
  \label{fig:lq22}
\end{figure}

\begin{figure}[H]
  \centering
  \begin{tabular}{@{}c@{}}
\minipage{0.3\textwidth} 
\center 
  \includegraphics[width=0.6\linewidth]{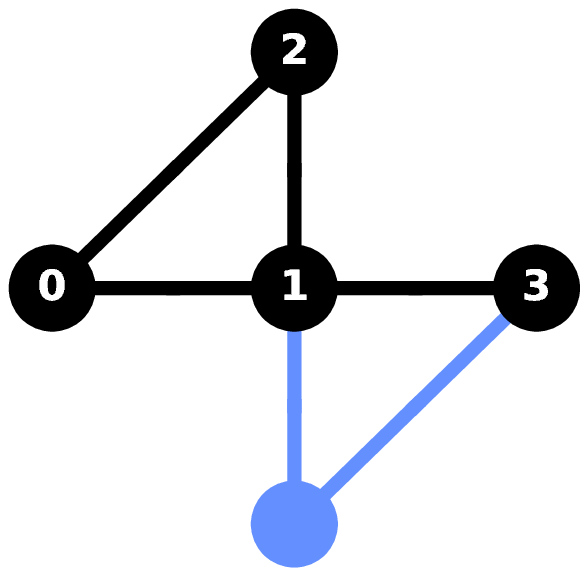} 
     \caption*{ibmqx2} 
\endminipage\hfill
\minipage{0.3\textwidth}
\center
  \includegraphics[width=0.6\linewidth]{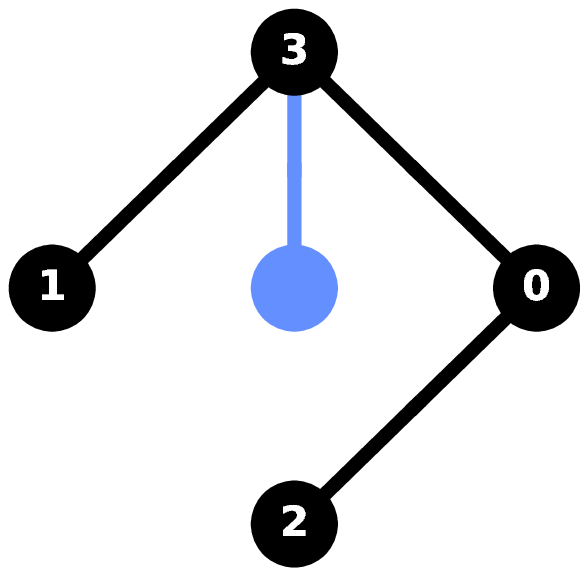} 
     \caption*{ibmq\_london}
 \endminipage\hfill
\minipage{0.3\textwidth}
\center
  \includegraphics[width=1\linewidth]{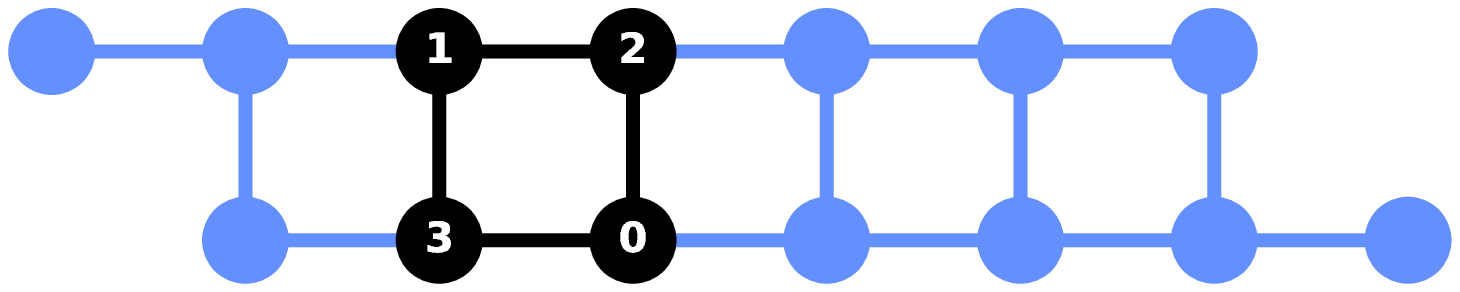} 
     \caption*{ibmq\_16\_melbourne}
 \endminipage   
  \end{tabular}
  \caption{QFT based shift, $8$ sites, $1$ step.}
  \label{fig:lq31}
\end{figure}

\newpage
\subsection{Smallest transpiled circuit sizes}  \label{subsec:sizetrans}

The  smallest circuit sizes that were obtained during simulations are given below for the respective experiments. These are not always the same as the circuits that gave the least $\ell^1$ distance from the ideal distribution  shown in the main body of the paper.

\begin{table}   [H]
\begin{center}
\begin{tabu}to\linewidth{|[1.5pt]c|c|[1.5pt]}  %
\tabucline[1.5pt]-
& (size , depth)\\
\tabucline[1.5pt]-
ibmqx2  &($11$, $7$)\\
\hline 
ibmq\_london & ($21$, $15$) \\
\hline 
ibmq\_$16$\_melbourne &($25$, $16$)\\ 
\tabucline[1.5pt]-
\end{tabu}
\end{center}
\caption{QFT based shift, $4$ sites, $1$ step}
 \label{table:q21}  
\end{table} 

\begin{table}   [H]
\begin{center}
\begin{tabu}to\linewidth{|[1.5pt]c|c|[1.5pt]}  %
\tabucline[1.5pt]-
& (size , depth)\\
\tabucline[1.5pt]-
ibmqx2 &($18$, $11$)\\
\hline 
ibmq\_london & ($38$, $29$) \\
\hline 
ibmq\_$16$\_melbourne &($48$, $30$)\\ 
\tabucline[1.5pt]-
\end{tabu}
\end{center}
\caption{QFT based shift, $4$ sites, $2$ steps}
 \label{table:q22}  
\end{table}

\begin{table}   [H]
\begin{center}
\begin{tabu}to\linewidth{|[1.5pt]c|c|[1.5pt]}  %
\tabucline[1.5pt]-
& (size , depth)\\
\tabucline[1.5pt]-
ibmqx2 &($47$, $33$)\\
\hline 
ibmq\_london &($59$, $38$) \\
\hline 
ibmq\_$16$\_melbourne &($78$, $45$)\\ 
\tabucline[1.5pt]-
\end{tabu}
\end{center}
\caption{QFT based shift, $8$ sites, $1$ step}
 \label{table:q31}  
\end{table} 

\end{document}